\documentclass[pre,twocolumn,amsmath,amssymb,showpacs,floatfix]{revtex4-1}
\usepackage{color,graphicx,amssymb,latexsym,mathrsfs,amsmath}
\usepackage{ulem,comment}
\begin{document}
\newcommand{\gae}{\lower 2pt \hbox{$\, \buildrel {\scriptstyle >}\over {\scriptstyle
\sim}\,$}}
\newcommand{\lae}{\lower 2pt \hbox{$\, \buildrel {\scriptstyle <}\over {\scriptstyle
\sim}\,$}}

\newcommand{\red}[1]{\textcolor{red}{#1}}
\newcommand{\blue}[1]{\textcolor{blue}{#1}}
\newcommand{\ave}[1]{\langle #1 \rangle}

\newcommand{\bS}{\boldsymbol{S}}
\newcommand{\bchi}{\boldsymbol{\chi}}
\newcommand{\bbm}{\boldsymbol{m}}
\newcommand{\bh}{\boldsymbol{h}}
\newcommand{\dfracp}[2]{\dfrac{\partial #1}{\partial #2}}
\newcommand{\wt}{\widetilde}
\newcommand{\wh}{\widehat}
\newcommand{\bzero}{\boldsymbol{0}}

\title{Critical exponents in mean-field classical spin systems}
\author{Yoshiyuki Y. Yamaguchi}
\affiliation{\mbox{Department of Applied Mathematics and Physics, Graduate
    School of Informatics, Kyoto University, Kyoto 606-8501, Japan}}
\author{Debraj Das}
\email[Corresponding author: ]{debraj.das@rkmvu.ac.in}
\affiliation{\mbox{Department of Physics, Ramakrishna Mission
    Vivekananda University, Belur Math, Howrah, 711202, India}}
\author{Shamik Gupta}
\affiliation{\mbox{Department of Physics, Ramakrishna Mission
    Vivekananda University, Belur Math, Howrah, 711202, India}}
\begin{abstract}
For mean-field classical spin systems
exhibiting a second-order phase transition in the stationary state, we
obtain within the corresponding phase space evolution according to the
Vlasov equation the values of the critical exponents describing
power-law behavior of response to a small external field. The
exponent values so obtained significantly differ from the ones obtained
on the basis of an analysis of the static phase-space distribution, with
no reference to dynamics.
This work serves as an illustration that cautions against relying on a static
approach, with no reference to the dynamical
evolution, to extract critical exponent values for mean-field systems.
\end{abstract}
\date{\today}
\maketitle

%%%%%%%%%%%%%%%%%%%%%%%%%%%%%%%%%%%%%%%%%%%%%%%%%%%%%%%%%%%%%%%%%%%%%%%%%%%%%%%%%%%%%%%%%%%
\section{Introduction}
\label{sec:intro}

Since early days of statistical mechanics, studying phase transitions in
physical systems has been a theme of active research in the field.
Phase transitions can occur only in the thermodynamic limit. Second-order or continuous phase
transitions are characterized by a power-law behavior of macroscopic quantities close to the critical point
of transition. Such transitions in different systems may be broadly
classified into universality classes identified by different values of critical exponents describing the
power-law behavior. For example, for a ferromagnet exhibiting a second-order
phase transition as a function of temperature $T$,
the magnetization close to and on the lower side of the critical point
$T_c$ has a power-law dependence on the separation $(T_c-T)$ from the critical
point, with the corresponding exponent being $\beta$. 
On applying an external field, the magnetization increases as a function
of the field strength, and in the limit of an infinitesimal field, a
linear growth for $T \ne T_c$ implying a linear response determines the zero-field susceptibility $\chi$.
The susceptibility diverges as a power law close to and on both sides of
the critical point, with the corresponding exponents denoted by $\gamma^{+}$ and $\gamma^{-}$
on the disordered ($T>T_c$) and the magnetized ($T<T_c$) phase,
respectively. At the critical point, the response becomes nonlinear,
being characterized by the critical exponent $\delta$.
These critical exponents are known to satisfy the scaling relation
$\gamma^{\pm}=\beta(\delta-1)$ \cite{Fisher,Stanley,nishimori-ortiz-11}.

One representative class of systems exhibiting second-order phase
transitions is that of mean-field systems. In thermal equilibrium of such
systems, statistical mechanical predictions for the critical exponents,
based on an analysis of the thermal equilibrium phase space
distribution with no reference to dynamics, yield the values
$\beta=1/2,~\gamma^{\pm}=1$, $\delta=3$~\cite{Stanley}.
However, owing to the mean-field nature of the time evolution, critical
exponents obtained on the basis of dynamics may well have different
values.
Indeed, dynamics of a mean-field system in the thermodynamic limit is
described by the so-called Vlasov equation that allows a vast number of
stable stationary states, and thermal equilibrium is just one of them~\cite{campa-dauxois-ruffo-09,levin-etal-14,Campa:2014,Gupta:2017}.
This implies that once the system is in a stable stationary state other
than thermal equilibrium, it would not relax to thermal equilibrium.
A large but finite system remains trapped in so-called quasistationary
states (QSSs) identified as stable stationary solutions of the Vlasov
equation, with finite-size effects allowing a slow evolution of the QSSs towards
thermal equilibrium over a timescale that diverges with the system size~\cite{yamaguchi-etal-04,binney-tremaine-08}.

  Existence of QSSs allows nonequilibrium phase transitions:
  a generic initial state undergoes a violent relaxation to relax to a
  QSS \cite{barre-etal-06}, and the nonequilibrium phase transition can
  for example be defined
  with respect to the value of the order parameter in the QSS.
  In a given system, these nonequilibrium phase transitions may not necessarily be continuous
  even when the equilibrium phase transition is continuous,
  and several discontinuous nonequilibrium phase transitions
  have been reported in the literature \cite{antoniazzi-07,Pakter:2011,rochafilho-amato-figueiredo-12,teles-benetti-pakter-levin-12}.
  In this article, we, however, focus on families of QSSs
  that exhibit continuous phase transitions
  and to which an external field is applied
  in order to investigate the values of the critical exponents
  characterizing the response.

The aforementioned trapping scenario holds even when an external field
is applied to the system prepared in a thermal equilibrium state: With
the field on, a finite system goes from the initial to a new thermal
equilibrium state via intermediate QSSs, while a thermodynamic system
remains trapped in a QSS and does not relax to thermal equilibrium~\cite{ogawa-patelli-yamaguchi-14}. The latter fact requires
that one invokes an alternative strategy of obtaining susceptibility
that is based on the Vlasov dynamics when addressing the issue of
response of mean-field systems in thermal equilibrium to an external
field.
The critical exponents $\gamma^{\pm}$ so obtained may not necessarily coincide
with the ones computed within equilibrium statistical mechanics.
Indeed, in the so-called Hamiltonian mean-field (HMF) model
\cite{inagaki-konishi-93,antoni-ruffo-95}, a paradigmatic mean-field
system exhibiting a second-order phase transition,
the critical exponents obtained within the Vlasov dynamics have been
shown to be  $\gamma^{+}=1,~\gamma^{-1}=1/4$
for a family of stable stationary initial states with $\beta=1/2$
\cite{ogawa-patelli-yamaguchi-14}.
Moreover, at the critical point, the Vlasov dynamics gives $\delta=3/2$.
More generally, the critical exponents within the Vlasov dynamics
have been obtained as
$\gamma^{+}=2\beta,~\gamma^{-}=\beta/2,~\delta=3/2$
for a class of Hamiltonian particle systems including the HMF model
\cite{ogawa-yamaguchi-15}.
Interestingly, the critical exponents obtained within the two approaches
satisfy the same scaling relation, namely, $\gamma^{-}=\beta(\delta-1)$.

  The difference in the values of the critical exponents
  that are obtained based on statistical mechanics and dynamics
  stems from the existence of an infinite number of so-called Casimir
  invariants that are constants of motion for the Vlasov dynamics
  and which make the dynamics non-ergodic.
  In general, existence of constraints suppresses susceptibility 
  \cite{mazur-69,suzuki-71}
  and accordingly the values of the critical exponents $\gamma^{\pm}$.
  This fact is borne out by the values of the critical exponents
  obtained in the HMF model by taking into account the Casimir constraints.
  Another important remark is that divergence of susceptibility
  is observed even when the dynamics is non-ergodic, as is found in the case
  of the HMF model \cite{ogawa-patelli-yamaguchi-14}.

The HMF model mimics the classical $XY$ model, with an additional kinetic
energy term assigned to individual spins. Owing to the latter whose range is the whole real set,
the one-particle phase space of the HMF model is a cylinder. In the HMF
model, the Poisson bracket between the spin components is taken to
vanish identically, 
In this work, we consider Heisenberg spin systems with mean-field interactions,
in which the Poisson brackets between the spin components are strictly
nonzero, and the single-particle phase space is the unit sphere.
Considering the time evolution of the spin components according to a
Hamiltonian with a mean-field interaction and a local
anisotropy, we address here several questions of theoretical and
practical relevance: Does the universality class for usual Hamiltonian
systems defined on a cylinder, e.g., the HMF model, 
include spin systems defined on the unit sphere? What is the effect of
the anisotropy on the critical exponents?
Would the scaling relation $\gamma^{-}=\beta(\delta-1)$ still hold even if the spin system
is found to be in a different universality class?

This paper is organized as follows.
The spin model we study is introduced in Sec. \ref{sec:model}. Here, the dynamics described by the canonical equations of motion
is also discussed, as is the characterization of the dynamics in the
thermodynamic limit in terms of the Vlasov equation. Based on the
latter, we discuss the setting and the definition of the critical
exponents in Sec. \ref{sec:setting}, while our theoretical predictions
for the critical exponents
are derived in Sec. \ref{sec:theory}. Detailed numerical checks of our
theoretical predictions are pursued in Sec. \ref{sec:numerics}.
Section \ref{sec:conclusions} concludes the paper with discussions.

%%%%%%%%%%%%%%%%%%%%%%%%%%%%%%%%%%%%%%%%%%%%%%%%%%%%%%%%%%%%%%%%%%%%%%%%%%%%%%%%%%%%%%%%%%%
\section{The model}
\label{sec:model}

%%%%%%%%%%%%%%%%%%%%%%%%%%%%%%%%%%%%%%%%%%%%%%%%%%%%%%%%%%%%%%%%%%%%%%%%%%%%%%%%%%%%%%%%%%%
%%%%%%%%%%%%%%%%%%%%%%%%%%%%%%%%%%%%%%%%%%%%%%%%%%%%%%%%%%%%%%%%%%%%%%%%%%%%%%%%%%%%%%%%%%%
\subsection{Definition}

Our model of study consists of $N$ globally-coupled classical Heisenberg spins
of unit length denoted by
\begin{equation}
  \bS_i=(S_{ix},S_{iy},S_{iz});\quad i=1,2,\ldots,N.
\end{equation}
The $N$-body Hamiltonian of the model is given by
\begin{equation}
  \label{eq:H}
  H_{N} = -\frac{J}{2N}\sum_{i,j=1}^N \bS_{i}\cdot\bS_{j}
  + D\sum_{i=1}^N S^{2n}_{iz}
  - \bh(t)\cdot \sum_{i=1}^{N} \bS_{i}.
\end{equation}
Here, the first term with $J > 0$ on the right-hand side models a ferromagnetic mean-field interaction between the spins.
The coupling constant $J$ 
has been scaled down by the system size $N$ in order to make the
energy extensive, in accordance with the Kac
prescription \cite{Kac:1963}.
The system~\eqref{eq:H} is however intrinsically non-additive:
it cannot be trivially subdivided into independent macroscopic parts.
In the following, we set $J=1$ without loss of generality.

In Eq.~\eqref{eq:H}, the second term with $D>0$ on the right-hand side accounts for
local anisotropy; Restricting to the subclass of models that are symmetric under $S_{iz}\to -S_{iz}$,
we have made here the choice of even exponent equal to $2n$, with $n$
being a non-negative integer.
We refer to the model with exponent $2n$ as Model-$n$. Note that
Model-$0$ is completely isotropic in the spin space, and there is no
preferred direction of orientation of spins.
Model-$1$ has been studied previously in the context of QSSs in
Refs.~\cite{Gupta:2011,Barre:2014}.
Model-$2$ is the special case of a quartic anisotropy; it may be
noted that thermodynamic properties of a Heisenberg spin model
containing a quartic term have been studied in Ref.~\cite{kventsel-katriel-84}.

The third term on the right-hand side of Eq.~\eqref{eq:H} arises due to the application
of a time-dependent external magnetic field
$\bh(t) \equiv (h_{x}(t),h_{y}(t),h_{z}(t))$. In this work, we consider
the external field to be absent for times previous to instant $t_0$, i.e., for times $t<t_0$, when the system will be
assumed to be existing in a reference state, e.g., a thermal equilibrium
state. For times $t \ge t_0$, on the other hand, we would put on a constant
field in order to measure the response of the reference state to the
external field. The
explicit form of $\bh(t)$ is thus given by 
\begin{equation}
  \bh(t) = \Theta(t-t_0) \bh,
\end{equation}
where $\Theta(t)$ is the unit step function, and $\bh$ is a vector of
constant length equal to $h$.
The singularity of the unit step function $\Theta(t)$
  will have no effect on the values of the critical exponents obtained based on the Vlasov dynamics,
  and we may replace $\Theta(t)$ with a smooth function
  \cite{ogawa-yamaguchi-12}.

%%%%%%%%%%%%%%%%%%%%%%%%%%%%%%%%%%%%%%%%%%%%%%%%%%%%%%%%%%%%%%%%%%%%%%%%%%%%%%%%%%%%%%%%%%%
%%%%%%%%%%%%%%%%%%%%%%%%%%%%%%%%%%%%%%%%%%%%%%%%%%%%%%%%%%%%%%%%%%%%%%%%%%%%%%%%%%%%%%%%%%%
\subsection{Spin dynamics}

In dimensionless times, the time evolution of system \eqref{eq:H} is governed by the set of
coupled first-order differential equations 
\begin{equation}
  \dot{\bS}_i=\{\bS_i,H_{N}\}; ~~~~i=1,2,\ldots,N,
\label{eq:eom}
\end{equation}
where the dot denotes derivative with respect to time.
The Poisson bracket $\{\cdot,\cdot\}$ is 
bilinear, skew-symmetric, and satisfies the Leibniz's rule
\begin{equation}
  \{XY, Z\} = \{X,Z\}Y + X\{Y,Z\}
\end{equation}
for any functions $X,Y,$ and $Z$ of the spins. 
The Poisson brackets between two spins are given by
\begin{equation}
  \begin{split}
    & \{S_{ix}, S_{jy}\} = \delta_{ij} S_{iz}, \\
    & \{S_{iy}, S_{jz}\} = \delta_{ij} S_{ix}, \\ 
    & \{S_{iz}, S_{jx}\} = \delta_{ij} S_{iy}.
  \end{split}
  \label{eq:spin-poisson-bracket}
\end{equation}
Using Eqs.~\eqref{eq:H}, \eqref{eq:eom}, and \eqref{eq:spin-poisson-bracket},
we obtain the time evolution of the spin components as
\begin{equation}
  \label{eq:eom-Si}
  \begin{split}
    \dot{S}_{ix} & = S_{iy}(m_z+h_{z})-S_{iz}(m_y+h_{y}) - 2nDS_{iy}S^{2n-1}_{iz}, \\
    \dot{S}_{iy} & = S_{iz}(m_x+h_{x})-S_{ix}(m_z+h_{z}) + 2nDS_{ix}S^{2n-1}_{iz}, \\
    \dot{S}_{iz} & = S_{ix}(m_y+h_{y})-S_{iy}(m_x+h_{x}),
  \end{split}
\end{equation}
where 
\begin{equation}
  \bbm \equiv \frac{1}{N} \sum_{i=1}^N \bS_i = (m_x,m_y,m_z)
\end{equation}
is the magnetization vector that serves as the mean field governing the
time evolution of the individual spins.
Summing the third equation of \eqref{eq:eom-Si} over $i$,
we find that $m_z$ is a constant of motion if the condition 
\begin{equation}
  m_{x} h_{y} - m_{y} h_{x} = 0
\end{equation}
is satisfied.
The length of each spin is a constant of motion, and so is the total
energy of the system when the field $\bh$ is time independent.

Writing the spin components in terms of spherical polar angles $\theta_i
\in [0,\pi]$ and $\phi_i \in [0,2\pi)$, as
\begin{equation}
  S_{ix} = \sin \theta_i \cos \phi_i,~
  S_{iy}=\sin \theta_i \sin \phi_i,~
  S_{iz}=\cos \theta_i,
\end{equation}
we obtain from Eq.~\eqref{eq:eom-Si} the time evolution of the variables $\theta_i$ and $\phi_i$ as
\begin{equation}
  \label{eq:eom-theta-phi}
  \begin{split}
    \dot{\theta}_i & = (m_x+h_{x})\sin\phi_i - (m_y+h_{y}) \cos\phi_i, \\
    \dot{\phi}_i & = (m_x+h_{x}) \cot\theta_i \cos\phi_i
    + (m_y+h_{y}) \cot\theta_i \sin\phi_i \\
    & - (m_z+h_{z}) + 2nD\cos^{2n-1}\theta_i. 
  \end{split}
\end{equation}

For later convenience, we introduce a new variable
$p_{i} \equiv \cos\theta_{i}$, in terms of which we have
\begin{equation}
  S_{ix} = \sqrt{1-p_{i}^{2}}\cos\phi_{i},~
  S_{iy} = \sqrt{1-p_{i}^{2}}\sin\phi_{i},~
  S_{iz} = p_{i}.
\end{equation}
In terms of $p_{i}$, which is in fact canonically
conjugate to $\phi_{i}$, the Poisson bracket $\{\cdot,\cdot\}$ reads \cite{Gupta:2011}
\begin{equation}
  \{ X, Y \}
  = \sum_{i=1}^{N} \left(
    \dfrac{\partial X}{\partial\phi_{i}}
    \dfrac{\partial Y}{\partial p_{i}}
    - \dfrac{\partial X}{\partial p_{i}}
    \dfrac{\partial Y}{\partial\phi_{i}}
  \right).
\end{equation}
The dynamical variables of the $i$-th spin are thus $\phi_i$ and $p_i$,
while a volume element in the $(\phi_{i},p_{i})$-space is ${\rm
d}\phi_{i}{\rm d}p_{i}$.

%%%%%%%%%%%%%%%%%%%%%%%%%%%%%%%%%%%%%%%%%%%%%%%%%%%%%%%%%%%%%%%%%%%%%%%%%%%%%%%%%%%%%%%%%%%
%%%%%%%%%%%%%%%%%%%%%%%%%%%%%%%%%%%%%%%%%%%%%%%%%%%%%%%%%%%%%%%%%%%%%%%%%%%%%%%%%%%%%%%%%%%
\subsection{Description in the thermodynamic limit}
\label{sec:large-population-limit}

In the thermodynamic limit $N\to\infty$, the dynamics of system \eqref{eq:H} is described by the Vlasov equation
\begin{equation}
  \label{eq:Vlasov}
  \dfracp{f}{t}
  + \dfracp{H}{p} \dfracp{f}{\phi} - \dfracp{H}{\phi} \dfracp{f}{p} = 0,
\end{equation}
where $f(\phi,p,t)$ is the single-spin distribution function that
measures the probability density to find a spin $(\phi,p)$ at time $t$, while 
the single-spin Hamiltonian $H$ is 
\begin{equation}
  \label{eq:single-H}
  H(\phi,p,t)
  = Dp^{2n} - [\bbm(t) +\bh(t)] \cdot \bS,
\end{equation}
with 
\begin{equation}
  \bS \equiv \left( \sqrt{1-p^{2}} \cos\phi, \sqrt{1-p^{2}} \sin\phi, p
  \right),
\end{equation}
and the magnetization vector $\bbm=(m_{x},m_{y},m_{y})$ given by
\begin{equation}
  \label{eq:mxmymz}
  \bbm(t)= \iint_{\mu} \bS f(\phi,p,t) {\rm d}\phi {\rm d}p.
\end{equation}
The double integral over any function $X(\phi,p)$ in the single-spin phase space $\mu \equiv
(\phi,p)$ is defined as
\begin{equation}
  \iint_{\mu} X(\phi,p) {\rm d}\phi {\rm d}p 
  \equiv \int_{0}^{2\pi} {\rm d}\phi \int_{-1}^{1} {\rm d}p~ X(\phi,p).
\end{equation}
Note that the single-spin Hamiltonian~\eqref{eq:single-H} depends on time $t$ through the magnetization
$\bbm(t)$ and the external field $\bh(t)$. Normalization of
$f(\phi,p,t)$ reads $\iint_\mu f(\phi,p,t)=1$ for any time $t$.

Any quantity
\begin{equation}
  C[f](t) = \iint_{\mu} c(f) {\rm d}\phi {\rm d}p
\end{equation}
is a constant of motion for any smooth function $c$, as may be seen by
considering the time variation of $C$ and using Eq.~\eqref{eq:Vlasov}.
These invariants of motion are called Casimir invariants,
which hold even when the single-spin Hamiltonian depends on time.
The Casimir invariants do not allow an initial state with $\bh=\bzero$ to relax to the thermal equilibrium state with $\bh\neq\bzero$
when at least one of the Casimir invariants $C[f]$ between the two states is not
the same.

%%%%%%%%%%%%%%%%%%%%%%%%%%%%%%%%%%%%%%%%%%%%%%%%%%%%%%%%%%%%%%%%%%%%%%%%%%%%%%%%%%%%%%%%%%%
\section{Setting and definition of the critical exponents}
\label{sec:setting}

%%%%%%%%%%%%%%%%%%%%%%%%%%%%%%%%%%%%%%%%%%%%%%%%%%%%%%%%%%%%%%%%%%%%%%%%%%%%%%%%%%%%%%%%%%%
%%%%%%%%%%%%%%%%%%%%%%%%%%%%%%%%%%%%%%%%%%%%%%%%%%%%%%%%%%%%%%%%%%%%%%%%%%%%%%%%%%%%%%%%%%%
\subsection{Setting}

For $t<t_0$, we consider system \eqref{eq:H} to be existing in one of a family of stable stationary states
with external field $\bh=\bzero$. In order that we may study the
critical exponents associated with the response of the system to an
external field that we put on for times $t\ge t_0$, we restrict to a
family of states that allow a second-order phase transition and
consequently a critical point in the stationary state. We refer to such
a family of states as our reference states and denote the states by
$f_{0}$. From Eq.~\eqref{eq:Vlasov}, it is evident that $f_{0}$ of the form
\begin{equation}
\label{eq:f0}
  f_{0}(\phi,p) = F(H_{0}(\phi,p))
  = \dfrac{G(H_{0}(\phi,p))}{\iint_{\mu} G(H_{0}(\phi,p)) {\rm d}\phi
  {\rm d}p},
\end{equation}
with $G$ an arbitrary function, is a stationary solution of the Vlasov equation, and we have
\begin{equation}
\label{eq:H0}
  H_{0}(\phi,p) = Dp^{2n} - \bbm_{0}\cdot\bS,
\end{equation}
and
\begin{equation}
  \bbm_{0} = (m_{0x},m_{0y},m_{0z})
\end{equation}
satisfying the self-consistent equation
\begin{equation}
  \bbm_{0} = \iint_{\mu} \bS f_{0}(\phi,p) {\rm d}\phi {\rm d}p.
  \label{eq:self-consistent}
\end{equation}

The family of functions $G$ may be parametrized by a parameter $T$,
which in the case of thermal equilibrium coincides with the
temperature~\cite{note}:
\begin{equation}
  \label{eq:Geq}
  G(x) = \exp(-x/T).
\end{equation}
However, the analysis presented in the following applies to other family
of functions $G$, such as the Fermi-Dirac-type family
\begin{equation}
  G(x) = \dfrac{1}{\exp[(x-a)/b]+1}.
\end{equation}
In this case, the parameter $T$ may be identified with either of the two
parameters $a$ and $b$.

Now, from the rotational symmetry of $H_0(\phi,p)$ on the
$(S_{x},S_{y})$-plane, we may set $m_{0y}=0$ without loss of generality.
Moreover, we may assume $m_{0z}=0$, which solves the self-consistent equation for $m_{0z}$.
Denoting $m_{0x}$ by $m_{0}$, so that $\bbm_{0}=(m_{0},0,0)$, we have 
\begin{equation}
  H_{0}(\phi,p) = Dp^{2n} - m_{0}\sqrt{1-p^{2}}\cos\phi,
\end{equation}
while the self-consistent equation~\eqref{eq:self-consistent} reads
\begin{equation}
  \label{eq:self-consistent-m0}
  m_{0} = \iint_{\mu} \sqrt{1-p^{2}}\cos\phi F(H_{0}(\phi,p)) {\rm
  d}\phi {\rm d}p.
\end{equation}

At $t=t_0$, we turn on a constant external field $\bh=(h,0,0)$ pointing in the direction of the reference magnetization
$\bbm_{0}=(m_{0},0,0)$. In presence of the external field, the system
evolving under the Vlasov dynamics~\eqref{eq:Vlasov} relaxes from the reference state $f_{0}$
to a stationary state $f_{h}$ with magnetization
$\bbm_{h}=(m_{h},0,0)$.
The single-spin Hamiltonian corresponding to the state $f_{h}$ is
\begin{equation}
  \label{eq:Hh}
  H_{h}(\phi,p) = Dp^{2n} - (m_{h}+h) \sqrt{1-p^{2}} \cos\phi.
\end{equation}
We stress that $f_{h}$ is not necessarily the thermal equilibrium state
proportional to $\exp[-H_{h}(\phi,p)/T]$, and thus could be an
out-of-equilibrium state, see Sec. \ref{sec:large-population-limit}.
Within the Vlasov dynamics, the response to the external field is measured by
\begin{equation}
  \delta m \equiv m_{h} - m_{0}.
\end{equation}
In the above setting, we recall the definitions of the
critical exponents $\beta,\gamma^{+},\gamma^{-}$ and $\delta$ given in any
standard reference on critical phenomena, e.g., Ref.~\cite{Fisher}.

%%%%%%%%%%%%%%%%%%%%%%%%%%%%%%%%%%%%%%%%%%%%%%%%%%%%%%%%%%%%%%%%%%%%%%%%%%%%%%%%%%%%%%%%%%%
%%%%%%%%%%%%%%%%%%%%%%%%%%%%%%%%%%%%%%%%%%%%%%%%%%%%%%%%%%%%%%%%%%%%%%%%%%%%%%%%%%%%%%%%%%%
\subsection{Definition of the critical exponents}

The critical exponent $\beta$ is defined with respect to the reference
state, as
\begin{equation}
  m_{0}(T) \propto (T_c-T)^{\beta};~~T \to T_c^-,
  \label{eq:beta}
\end{equation}
where $T_c$ is the critical point. Here, $m_{0}$ is the positive solution of the self-consistent
  equation \eqref{eq:self-consistent-m0},
  and the value of $\beta$ may depend on the choice of the family $F$ of the reference state.
  The self-consistent equation \eqref{eq:self-consistent-m0},
  however, implies quite generally that $\beta=1/2$, see Appendix
  \ref{sec:StatMech}.

The critical exponents $\gamma^{\pm}$ are defined in the regime of linear response.
The response $\delta m$ depends on $T$ and $h$, 
and the susceptibility $\chi(T)$ is defined as
\begin{equation}
  \chi(T) \equiv \left. \dfrac{\partial (\delta m)}{\partial h} \right|_{h\to 0}.
\end{equation}
The susceptibility diverges at the critical point $T_c$ as
\begin{equation}
  \chi(T) \propto \left\{
    \begin{array}{ll}
      (T-T_c)^{-\gamma^{+}} & (T \to T_c^+), \\
      (T_c-T)^{-\gamma^{-}} & (T \to T_c^-),\\
    \end{array}
    \label{eq:gamma}
  \right. 
\end{equation}
which defines the exponents $\gamma^\pm$.

At the critical point $T_c$, one has
\begin{equation}
  \delta m \propto h^{1/\delta}, \quad (T=T_c),
  \label{eq:delta}
\end{equation}
which defines the critical exponent $\delta$.
Usually, one has $\delta>1$, since the leading response
is nonlinear and is stronger than the linear response.

%%%%%%%%%%%%%%%%%%%%%%%%%%%%%%%%%%%%%%%%%%%%%%%%%%%%%%%%%%%%%%%%%%%%%%%%%%%%%%%%%%%%%%%%%%%
%%%%%%%%%%%%%%%%%%%%%%%%%%%%%%%%%%%%%%%%%%%%%%%%%%%%%%%%%%%%%%%%%%%%%%%%%%%%%%%%%%%%%%%%%%%
\subsection{Statistical mechanics predictions for the critical exponents}

Statistical mechanics analysis that considers studying 
the equivalent of Eqs.~\eqref{eq:f0}, \eqref{eq:H0} and
\eqref{eq:self-consistent} in presence of a constant field
$\bh=(h,0,0)$, with no reference to dynamics, gives
\begin{equation}
  \label{eq:exponents-SM}
  \beta = \dfrac{1}{2}, \quad
  \gamma^{\pm} = 1, \quad
  \delta = 3,
\end{equation}
irrespective of the value of the exponent $n$, see Appendix \ref{sec:StatMech} for details.
In the next section, we derive the values of the critical exponents
within the Vlasov dynamics.
We will obtain the response
$\delta m$ within the Vlasov dynamics, and hence the
exponents $\gamma^{\pm}$ and $\delta$ may very well take values
different from the ones in Eq.~\eqref{eq:exponents-SM}.

%%%%%%%%%%%%%%%%%%%%%%%%%%%%%%%%%%%%%%%%%%%%%%%%%%%%%%%%%%%%%%%%%%%%%%%%%%%%%%%%%%%%%%%%%%%
\section{Theoretical predictions for the critical exponents based on the
Vlasov dynamics}
\label{sec:theory}

In this section, we derive our results for the critical exponents based
on the Vlasov dynamics. 
As already mentioned above, we have $\beta=1/2$ quite generally
for all choices of the family $F$ of the reference state.
In the following, we discuss the
computation of the critical exponents $\gamma^{\pm}$ and $\delta$ for a
given value of the exponent $\beta$.

%%%%%%%%%%%%%%%%%%%%%%%%%%%%%%%%%%%%%%%%%%%%%%%%%%%%%%%%%%%%%%%%%%%%%%%%%%%%%%%%%%%%%%%%%%%
%%%%%%%%%%%%%%%%%%%%%%%%%%%%%%%%%%%%%%%%%%%%%%%%%%%%%%%%%%%%%%%%%%%%%%%%%%%%%%%%%%%%%%%%%%%
\subsection{Model-$0$}

The single-spin Hamiltonian of Model-$0$ is
\begin{equation}
  H_{h} = -(m_{h}+h) S_{x}.
\end{equation}
The equations of motion are obtained from Eq.~(\ref{eq:eom-Si}) as 
\begin{equation}
  \dot{S}_{ix} = 0,~\dot{S}_{iy} = S_{iz}(m_h+h),~\dot{S}_{iz} =
  -S_{iy}(m_h+h).
  \label{eq:model0}
\end{equation}
Clearly, $S_{ix}$ for any $i$ and consequently $m_{x}$ are
 constant of motion for any external field $h$
  irrespective of its time dependence.
  Dynamically, each spin rotates
on a $S_{x}=$ constant plane. The fact that the variable $S_{x}$ is a constant of motion
for both cases of $h=0$ and $h\neq 0$ implies that the reference state
$f_{0}=F(H_{0})=F(-m_{0}S_{x})$
is stationary even after the external field is turned on, 
and we have $m_{h}=m_{0}$.
Consequently, no response to the external field is obtained within the
Vlasov dynamics.
If we have to assign values to the critical exponents,
we may say
\begin{equation}
  \label{eq:exponents-Model-0}
  \beta=\dfrac{1}{2},\quad \gamma^{\pm}=0, \quad \delta=1, 
\end{equation}
owing to the fact that no divergence of the susceptibility is obtained
within the Vlasov dynamics. 
The aforementioned exponent values are quite different from the ones obtained within statistical mechanics, Eq.~\eqref{eq:exponents-SM}.
For the case $G(x)=\exp(-x/T)$, 
the critical point is $T_{\rm c}=1/3$ (obtained by using the results
in Appendix~\ref{sec:StatMech}, in particular, by substituting such a form of $G(x)$ into
the function $A(T)$ defined by Eq.~\eqref{eq:AT} and then solving
$A(T_{\rm c})=0$).

  The isotropic spin model, Model-$0$, shows no response to the external field
  and thus provides a simple and extreme example
  of dynamical suppression of response, but
  Model-$n$ with $n\geq 1$ does show non-zero response.
In the following subsection, we obtain the values of the critical exponents
for $n\geq 1$.

%%%%%%%%%%%%%%%%%%%%%%%%%%%%%%%%%%%%%%%%%%%%%%%%%%%%%%%%%%%%%%%%%%%%%%%%%%%%%%%%%%%%%%%%%%%
%%%%%%%%%%%%%%%%%%%%%%%%%%%%%%%%%%%%%%%%%%%%%%%%%%%%%%%%%%%%%%%%%%%%%%%%%%%%%%%%%%%%%%%%%%%
\subsection{Model-$n$ with $n \ge 1$}
\label{sec:response-formula}

To compute the values of the critical exponents $\gamma^{\pm}$ and $\delta$,
our task is to obtain within the Vlasov dynamics starting from the state
$f_0$ the asymptotic state $f_{h}$ and hence the response $\delta m$. For this purpose, we remark
that the Hamiltonian $H_{h}$ given in Eq.~\eqref{eq:Hh} is integrable and has the associated angle-action
variables $(w,I)$. In this setting, the response formula
\begin{equation}
  f_{h}(I) = \ave{f_{0}(\phi,p)}_{h}
\label{eq:response-formula}
\end{equation}
has been proposed for Hamiltonian systems
\cite{ogawa-yamaguchi-12,ogawa-yamaguchi-15},
where $\ave{\cdot}_{h}$ is defined as the average over the angle variable
$w$, as
\begin{equation}
  \ave{A}_{h} \equiv \dfrac{1}{2\pi} \int_{0}^{2\pi}
  A \left( \phi\left( w,I \right),
    p\left( w,I \right) \right) {\rm d}w.
\end{equation}
In Appendix~\ref{app:response}, we summarize the derivation of
Eq.~\eqref{eq:response-formula}.
Using $H_{h}=H_{h}(I)$ and
\begin{equation}
  \ave{\varphi(I)}_{h} = \varphi(I)
\end{equation}
for any function $\varphi$,
and the expansion of the single-spin Hamiltonian as
\begin{equation}
  H_{h} = H_{0} + \delta H,
  \quad
  \delta H = - (\delta m + h) S_{x},
\end{equation}
with $S_{x}=\sqrt{1-p^{2}} \cos\phi$,
we have the expansion of $f_{h}(I)$ as
\begin{equation}
  \label{eq:expansion-fV}
  \begin{split}
    f_{h}
    & = f_{0}
    - (\delta m+h) \left[ S_{x} - \ave{S_{x}}_{0} \right] F'(H_{0}) \\
    & - (\delta m+h) \left[ \ave{S_{x}}_{0} F'(H_{0})
    - \ave{S_{x}}_{h} F'(H_{h}) \right].
  \end{split}
\end{equation}
Note that, for instance, one has $F'(x)=-F(x)/T$
for thermal equilibrium reference state \eqref{eq:Geq}.
The average $\ave{\cdot}_{0}$ is defined as an average
over the angle variable associated with the integrable system $H_{0}$.

Multiplying Eq.~\eqref{eq:expansion-fV} by $S_{x}=\sqrt{1-p^{2}}\cos\phi$ 
and then integrating over $\phi$ and $p$, we have the self-consistent equation
for the response $\delta m$ as
\begin{equation}
  \label{eq:deltam}
  L (\delta m+h) + {\cal N}(\delta m+h) - h = \text{(higher order terms in $h$)},
\end{equation}
where the coefficient $L$ of the linear part is
\begin{equation}
  \label{eq:LT}
  L(T) = 1 + \iint_{\mu}
  \left[ S_{x}^{2} - \ave{S_{x}}_{0}^{2} \right] F'(H_{0}(\phi,p)) {\rm
  d}\phi {\rm d}p,
\end{equation}
while ${\cal N}$ concerns the leading nonlinear part:
\begin{equation}
  {\cal N}(T) = \iint_{\mu} \left[ \ave{S_{x}}_{0}^{2} F'(H_{0})
    - \ave{S_{x}}_{h}^{2} F'(H_{h}) \right] {\rm d}\phi {\rm d}p.
    \label{eq:NT}
\end{equation}
The linear part $L$ gives the values of the critical exponents $\gamma^{\pm}$, while the nonlinear part ${\cal N}$ gives the value of the exponent $\delta$.

Note that $L(T_c)=0$, so that the contribution of only the nonlinear response appears
at the critical point $T_c$.
Away from the critical point, it is the linear part that gives the
dominant contribution in Eq.~\eqref{eq:deltam},
so that neglecting the
nonlinear contribution, we have the linear response
\begin{equation}
  \delta m = \dfrac{1-L}{L} h.
  \label{eq:deltam-1}
\end{equation}
Then, within linear response, the divergence of the susceptibility,
\begin{equation}
  \label{eq:chi}
  \chi = \dfrac{1-L}{L},
\end{equation}
is determined by the convergence behavior of $L$
as $T\to T_c^{\pm}$.

%%%%%%%%%%%%%%%%%%%%%%%%%%%%%%%%%%%%%%%%%%%%%%%%%%%%%%%%%%%%%%%%%%%%%%%%%%%%%%%%%%%%%%%%%%%
%%%%%%%%%%%%%%%%%%%%%%%%%%%%%%%%%%%%%%%%%%%%%%%%%%%%%%%%%%%%%%%%%%%%%%%%%%%%%%%%%%%%%%%%%%%
%%%%%%%%%%%%%%%%%%%%%%%%%%%%%%%%%%%%%%%%%%%%%%%%%%%%%%%%%%%%%%%%%%%%%%%%%%%%%%%%%%%%%%%%%%%
\subsubsection{Linear response in the disordered phase}

In the disordered phase, the angle variable $w$ is nothing but $\phi$,
and we have
\begin{equation}
  \ave{S_{x}}_{0} = \ave{\sqrt{1-p^{2}}\cos\phi}_{0} = 0.
\end{equation}
This result implies that
$L$ has no contribution from the dynamics, and hence may be expanded in
a Taylor series in $(T-T_{\rm c})$ around $T_c$,
resulting in its convergence being proportional to $T-T_c$.
The critical exponent $\gamma^{+}$ is, therefore, given by
\begin{equation}
  \gamma^{+}=1.
\end{equation}

%%%%%%%%%%%%%%%%%%%%%%%%%%%%%%%%%%%%%%%%%%%%%%%%%%%%%%%%%%%%%%%%%%%%%%%%%%%%%%%%%%%%%%%%%%%
%%%%%%%%%%%%%%%%%%%%%%%%%%%%%%%%%%%%%%%%%%%%%%%%%%%%%%%%%%%%%%%%%%%%%%%%%%%%%%%%%%%%%%%%%%%
%%%%%%%%%%%%%%%%%%%%%%%%%%%%%%%%%%%%%%%%%%%%%%%%%%%%%%%%%%%%%%%%%%%%%%%%%%%%%%%%%%%%%%%%%%%
\subsubsection{Linear response in the ordered phase}

In the ordered phase, let us divide $L$ into the two parts:
\begin{equation}
  L(T) = L_{1}(T) + L_{2}(T),
\end{equation}
with
\begin{equation}
  L_{1}(T) \equiv 1 + \iint_{\mu} S_{x}^{2} F'(H_{0}(\phi,p)) {\rm
  d}\phi {\rm d}p
\end{equation}
and
\begin{equation}
  \label{eq:L2}
  L_{2}(T) \equiv - \iint_{\mu} \ave{S_{x}}_{0}^{2} F'(H_{0}(\phi,p))
  {\rm d}\phi {\rm d}p.
\end{equation}
The behavior of $L_1(T)$ is as in the disordered phase discussed above: $L_{1}(T)=O(T_c-T)$.
If $L_{2}(T)$ has slower convergence than $L_{1}(T)$,
the convergence of $L(T)$ will be dominated by that of $L_{2}(T)$.

In the HMF model, we can construct the angle-action variables explicitly,
and the estimation of $L_{2}(T)$ is rather straightforward.
In our spin model, such an explicit construction does not seem feasible
owing to the form of the single-spin Hamiltonian $H_{0}$, so that we have to invoke some physical observations and assumptions
in order to estimate $L_{2}(T)$.
Details of the estimation are presented in Appendix
\ref{sec:estimation-L2}, and one gets
\begin{equation}
  \label{eq:L2-estimation}
  L_{2}(T) = O((m_{0})^{1/(n+1)}) = O((T_c-T)^{\beta/(n+1)}).
\end{equation}
We remark that this estimation does not depend on
the choice of the reference family $G$.
From the above equation, it follows that the critical exponent $\gamma^{-}$ is
\begin{equation}
  \gamma^{-} = \dfrac{\beta}{n+1},
\end{equation}
provided $\beta\leq n+1$, which is satisfied for $\beta=1/2$ and $n\geq 1$.

%%%%%%%%%%%%%%%%%%%%%%%%%%%%%%%%%%%%%%%%%%%%%%%%%%%%%%%%%%%%%%%%%%%%%%%%%%%%%%%%%%%%%%%%%%%
%%%%%%%%%%%%%%%%%%%%%%%%%%%%%%%%%%%%%%%%%%%%%%%%%%%%%%%%%%%%%%%%%%%%%%%%%%%%%%%%%%%%%%%%%%%
%%%%%%%%%%%%%%%%%%%%%%%%%%%%%%%%%%%%%%%%%%%%%%%%%%%%%%%%%%%%%%%%%%%%%%%%%%%%%%%%%%%%%%%%%%%
\subsubsection{Nonlinear response at the critical point}

As mentioned earlier, $L(T_c)=0$, and Eq.~\eqref{eq:deltam} gives to leading order in
$h$ the result
\begin{equation}
  {\cal N}(T_c) (\delta m+h) - h = 0
\end{equation}
at the critical point $T_c$.
The first term of ${\cal N}(T)$, see Eq.~\eqref{eq:NT}, vanishes on
using the fact that $m_0=0$ at $T=T_c$ gives $\ave{S_{x}}_{0}=0$. As a
result, ${\cal N}(T_c)$ becomes
\begin{equation}
  {\cal N}(T_c)
  = - \iint_{\mu} \ave{S_{x}}_{h}^{2} F'(H_{h}(\phi,p)) {\rm d}\phi {\rm
  d}p,
\end{equation}
a form that reduces to the one for $L_{2}(T)$, Eq.~\eqref{eq:L2},
on replacing in the latter the reference state $f_{0}$
with the asymptotic state $f_{h}$ in performing the average over $S_x$.
We thus have an estimation of ${\cal N}(T_c)$ as
\begin{equation}
  {\cal N}(T_c) = O( (\delta m+h)^{1/(n+1)} ),
\end{equation}
where we have used Eq.~\eqref{eq:L2-estimation} and have replaced
$m_{0}$ in it with $m_{h}+h=\delta m+h$.
This estimation gives
\begin{equation}
  (\delta m+h)^{(n+2)/(n+1)} \propto h
\end{equation}
and hence, that
\begin{equation}
  \delta m \propto h^{(n+1)/(n+2)}.
\end{equation}
The critical exponent $\delta$ is thus
\begin{equation}
  \delta = \dfrac{n+2}{n+1}.
\end{equation}

%%%%%%%%%%%%%%%%%%%%%%%%%%%%%%%%%%%%%%%%%%%%%%%%%%%%%%%%%%%%%%%%%%%%%%%%%%%%%%%%%%%%%%%%%%%
%%%%%%%%%%%%%%%%%%%%%%%%%%%%%%%%%%%%%%%%%%%%%%%%%%%%%%%%%%%%%%%%%%%%%%%%%%%%%%%%%%%%%%%%%%%
%%%%%%%%%%%%%%%%%%%%%%%%%%%%%%%%%%%%%%%%%%%%%%%%%%%%%%%%%%%%%%%%%%%%%%%%%%%%%%%%%%%%%%%%%%%
\subsubsection{Predicted critical exponents and the scaling relation}

The theoretically predicted critical exponents, obtained within the
Vlasov dynamics, are thus
\begin{equation}
  \beta = \dfrac{1}{2},
  \quad
  \gamma^{+} = 1,
  \quad
  \gamma^{-} = \dfrac{\beta}{n+1},
  \quad
  \delta = \dfrac{n+2}{n+1},
  \quad
  (n\geq 1).
\end{equation}
These exponents satisfy the scaling relation
\begin{equation}
  \gamma^{-} = \beta (\delta-1),
\end{equation}
irrespective of the value of $\beta$.

We remark that Model-$0$ corresponds to the limit $n\to\infty$,
since this limit eliminates from the Hamiltonian the anisotropic term $Dp^{2n}$ for $|p|<1$.
In this limit, we have $\gamma^{-}=0$ and $\delta=1$,
which is consistent with \eqref{eq:exponents-Model-0} and
no response in Model-$0$.
The obtained critical exponents are displayed in Table \ref{tab:critical-exponents}.

\begin{table}
  \centering
  \begin{tabular}{|l|l|cccc|}
    \hline
    & Model-$n$ & $\beta$ & $\gamma^{+}$ & $\gamma^{-}$ & $\delta$ \\
    \hline \hline
    Statistical Mechanics & $n\geq 0$ & $1/2$ & $1$ & $1$ & $3$ \\
    \hline
    Vlasov dynamics & $n=0$ & $1/2$ & $0$ & $0$ & $1$ \\
    & $n\geq 1$ & $\dfrac{1}{2}$ & $1$ & $\dfrac{\beta}{n+1}$ & $\dfrac{n+2}{n+1}$ \\
    \hline
  \end{tabular}
  \caption{Critical exponents of the spin model \eqref{eq:H} obtained
  within the Vlasov dynamics. The critical exponents $\gamma^{\pm}$ and $\delta$ in Model-$0$
    reflect no response within the Vlasov dynamics.
    The scaling relation $\gamma^{-}=\beta(\delta-1)$ holds for all cases.}
  \label{tab:critical-exponents}
\end{table}

%%%%%%%%%%%%%%%%%%%%%%%%%%%%%%%%%%%%%%%%%%%%%%%%%%%%%%%%%%%%%%%%%%%%%%%%%%%%%%%%%%%%%%%%%%%
\section{Numerical tests}
\label{sec:numerics}

In this section, we discuss numerical checks of our theoretical
predictions for the critical exponents obtained in the preceding
section. As a representative case,
we focus on thermal equilibrium states as the reference states, which
are represented by Eq.~\eqref{eq:Geq} and give $\beta=1/2$.
We, however, underline that the theoretical results
developed in Sec.~\ref{sec:theory} hold
for other families of reference states.

  The numerical simulations of the equations of motion (\ref{eq:eom-Si})
  are performed by using a fourth-order Runge-Kutta algorithm
  with the timestep $\delta t=0.01$.
  We choose sufficiently large numbers of spins,
  namely $N=10^{6}$ or $10^{7}$.
  We will refer to $T$ as the temperature,
  but it is just a parameter characterizing the reference
  state~(\ref{eq:Geq}) and there is no thermal noise in the dynamics.

%%%%%%%%%%%%%%%%%%%%%%%%%%%%%%%%%%%%%%%%%%%%%%%%%%%%%%%%%%%%%%%%%%%%%%%%%%%%%%%%%%%%%%%%%%%
%%%%%%%%%%%%%%%%%%%%%%%%%%%%%%%%%%%%%%%%%%%%%%%%%%%%%%%%%%%%%%%%%%%%%%%%%%%%%%%%%%%%%%%%%%%
\subsection{Temporal evolution of magnetization}

\begin{figure}[ht!]
  \centering
  \includegraphics[width=7cm]{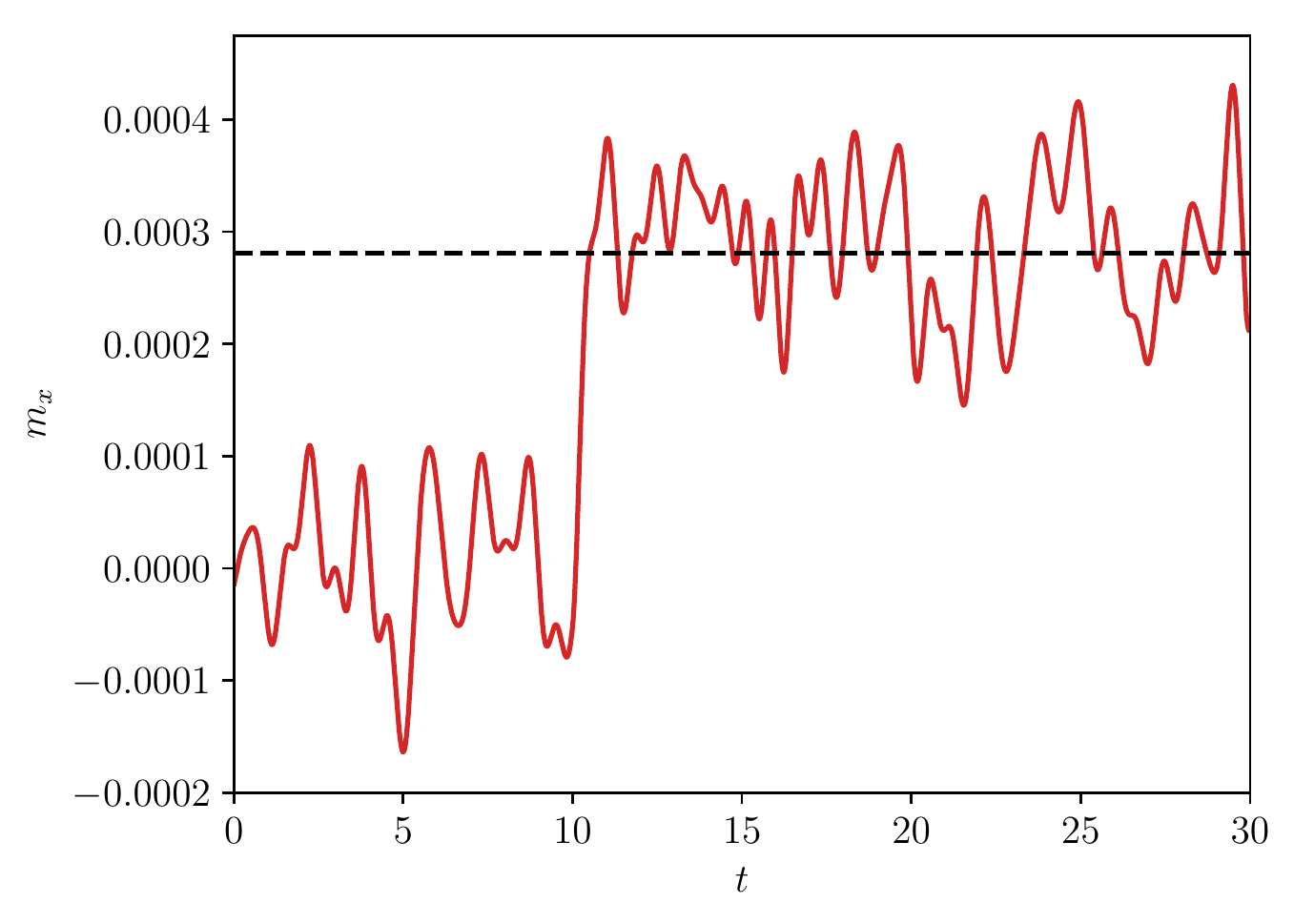}
  \caption{{\bf Model-$1$:} Considering $D=5$ and $N=10^6$, the figure
  shows the temporal evolution of $m_{x}$.
    The initial state is the thermal equilibrium state at temperature
    $T (>T_c \approx 0.476)=12.8$.
    A constant field of strength $h=0.01$ along the $x$-axis
    is turned on at time $t_0=10$. The dashed line gives the value of the
magnetization induced by the field and given by Eq.~(\ref{eq:deltam-1}) to be $\approx 0.00028$. The data are obtained by numerically integrating the equations of
  motion~\eqref{eq:eom-Si} and averaging over $100$ realizations of the dynamics.}  
\label{fig:n2-mag}
\end{figure}

  Considering $n=1$, and preparing the system in the thermal
  equilibrium state at a temperature $T>T_{\rm c}$,
  we show in Fig.~\ref{fig:n2-mag} the behavior of the magnetization $m_x$
  as a function of time when a constant field of strength $h=0.01$
  along the $x$-axis is turned on at $t_0=10$.
In the figure, we also show by the dashed line the value of the
magnetization induced by the field and given by Eq.~(\ref{eq:deltam-1}).
We can numerically examine the critical exponents
  $\gamma^{\pm}$ and $\delta$ by varying the temperature $T$
  and observing the response that corresponds to the difference between
  the zero level and the dashed-line level of the magnetization in the
  figure.

 The external field may have induced periodic oscillations in
the magnetization~\cite{Pakter:2013},
but in our case, one may observe from Fig.~\ref{fig:n2-mag} that the magnetization
does not exhibit stable oscillations
after the field is turned on.
It rather exhibits due to finiteness of the number of spins only fluctuations about the
theoretically predicted value valid in the thermodynamic limit, 
as has also been observed in the HMF model \cite{ogawa-patelli-yamaguchi-14}.

%%%%%%%%%%%%%%%%%%%%%%%%%%%%%%%%%%%%%%%%%%%%%%%%%%%%%%%%%%%%%%%%%%%%%%%%%%%%%%%%%%%%%%%%%%%
%%%%%%%%%%%%%%%%%%%%%%%%%%%%%%%%%%%%%%%%%%%%%%%%%%%%%%%%%%%%%%%%%%%%%%%%%%%%%%%%%%%%%%%%%%%
\subsection{Critical exponents}

\begin{figure}[ht!]
  \centering
  \includegraphics[width=7cm]{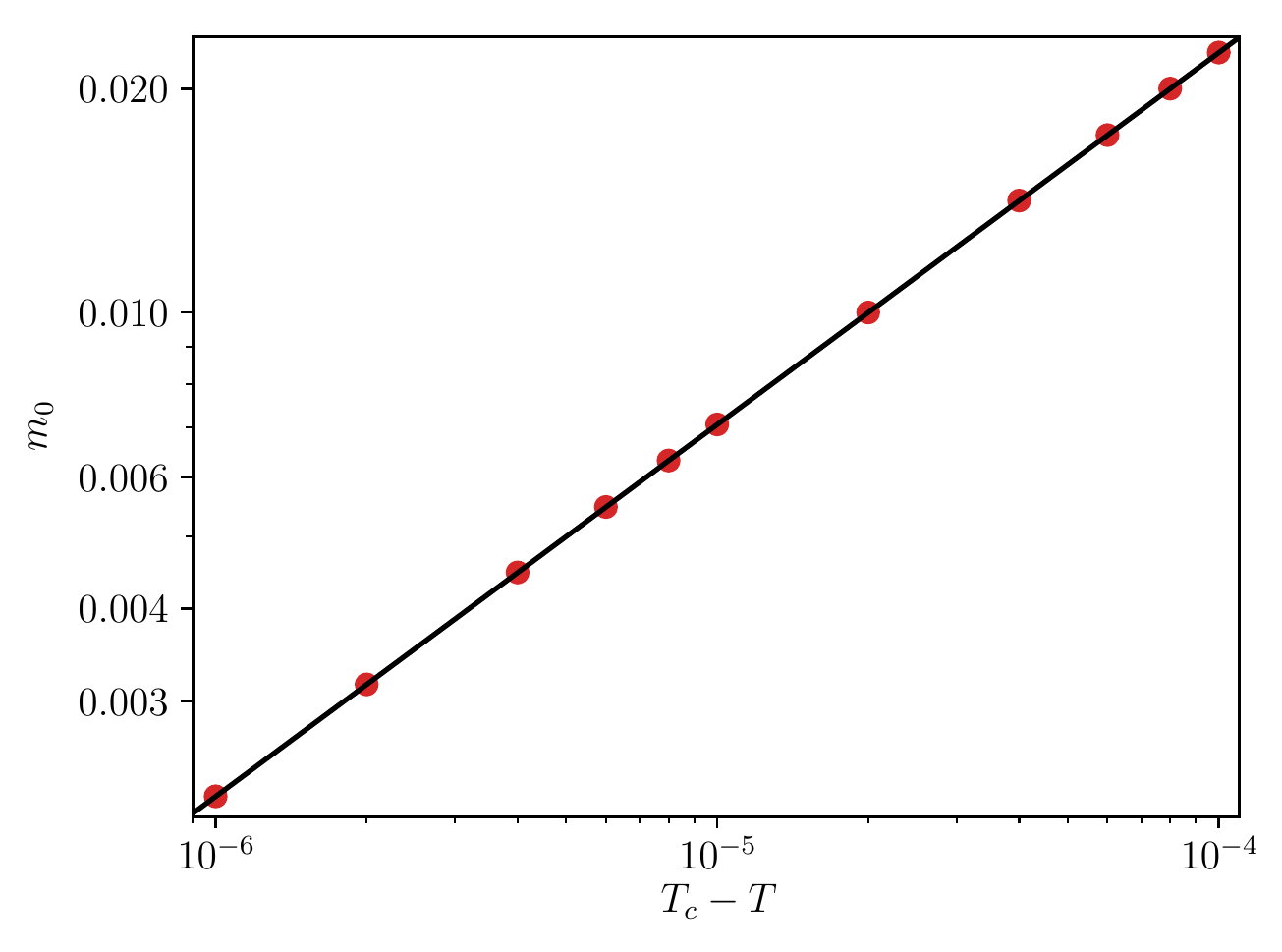}
   \caption{{\bf Model-$0$:} Spontaneous magnetization $m_{0}$ (points), obtained by
     solving the self-consistent equation~\eqref{eq:self-consistent-m0}
     with thermal equilibrium as the reference state. The critical point is $T_c=1/3$. The line corresponds to the behavior~\eqref{eq:beta}, with
  $\beta$ given by our theoretical analysis as $\beta=1/2$, see
  Table~\ref{tab:critical-exponents}.}
  \label{fig:n0-beta}
\end{figure}

\begin{figure}[ht!]
  \centering
  \includegraphics[width=7cm]{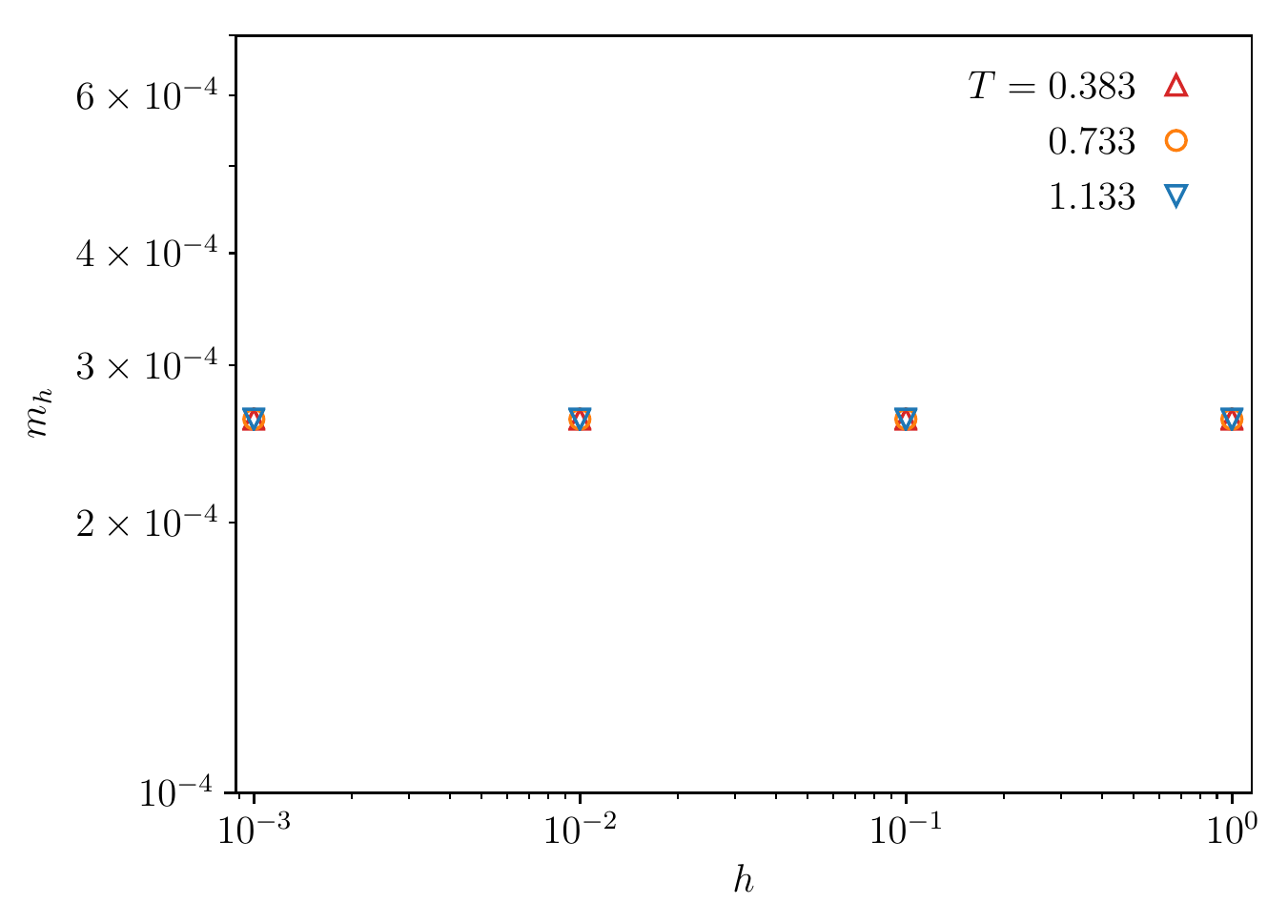}
  \includegraphics[width=7cm]{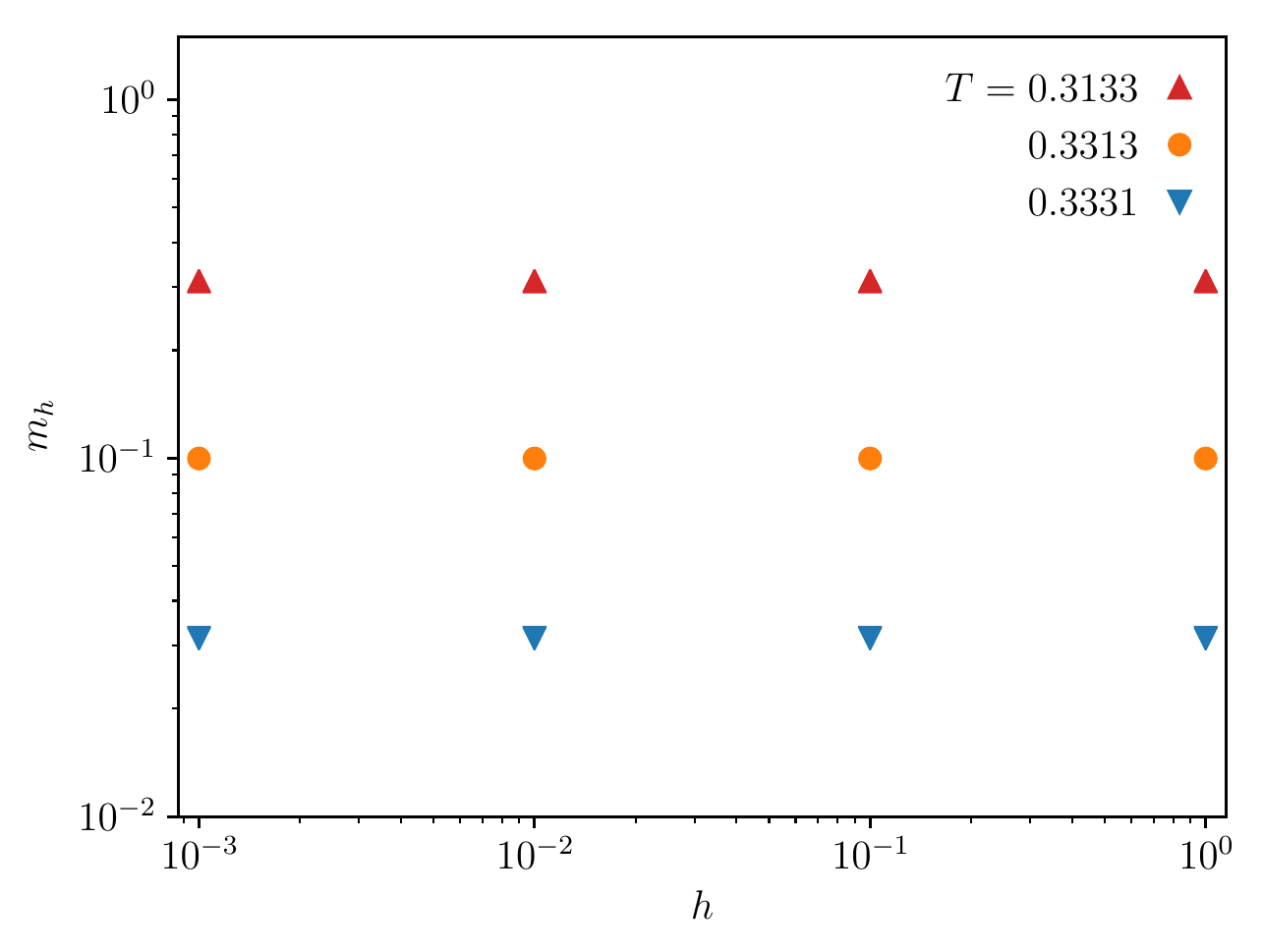}
  \caption{{\bf Model-$0$:} For $T>T_c=1/3$ (upper panel) and $T < T_c$
  (lower panel), the figure shows the magnetization
  $m_{h}$ obtained by numerically integrating the equations of
  motion~\eqref{eq:model0} with $N=10^7$ and performing a time average
  of the instantaneous magnetization over an interval of length
  $20$, which is further averaged over $5$ realizations of the dynamics.
  The magnetization $m_h$ is independent of $h$ in both cases, thus
  lending support to the behavior~\eqref{eq:gamma}, with
  $\gamma^\pm$ given by our theoretical analysis as $\gamma^+=\gamma^-=0$, see
  Table~\ref{tab:critical-exponents}.}
  \label{fig:n0-gamma}
\end{figure}

\begin{figure}[ht!]
  \centering
  \includegraphics[width=7cm]{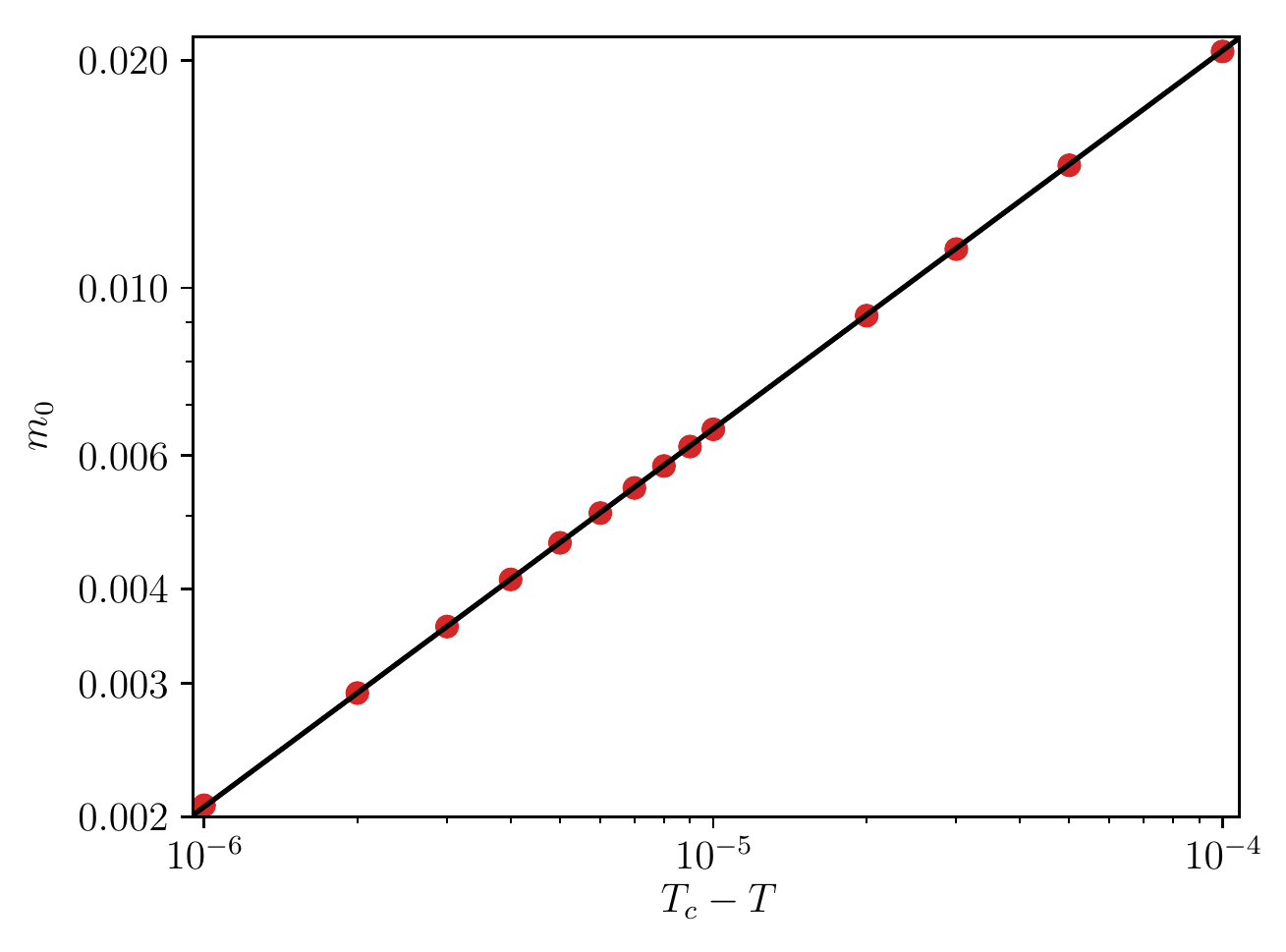}
  \caption{{\bf Model-$1$:} For $D=5$, the figure shows the spontaneous magnetization $m_{0}$ (points), obtained by
  solving the self-consistent equation \eqref{eq:self-consistent-m0}
  with thermal equilibrium as the reference state. The critical point is $T_c \approx 0.476$. The line corresponds to the behavior~\eqref{eq:beta}, with
  $\beta$ given by our theoretical analysis as $\beta=1/2$, see
  Table~\ref{tab:critical-exponents}.}
  \label{fig:n1-beta}
\end{figure}

\begin{figure}[ht!]
  \centering
 \includegraphics[width=7cm]{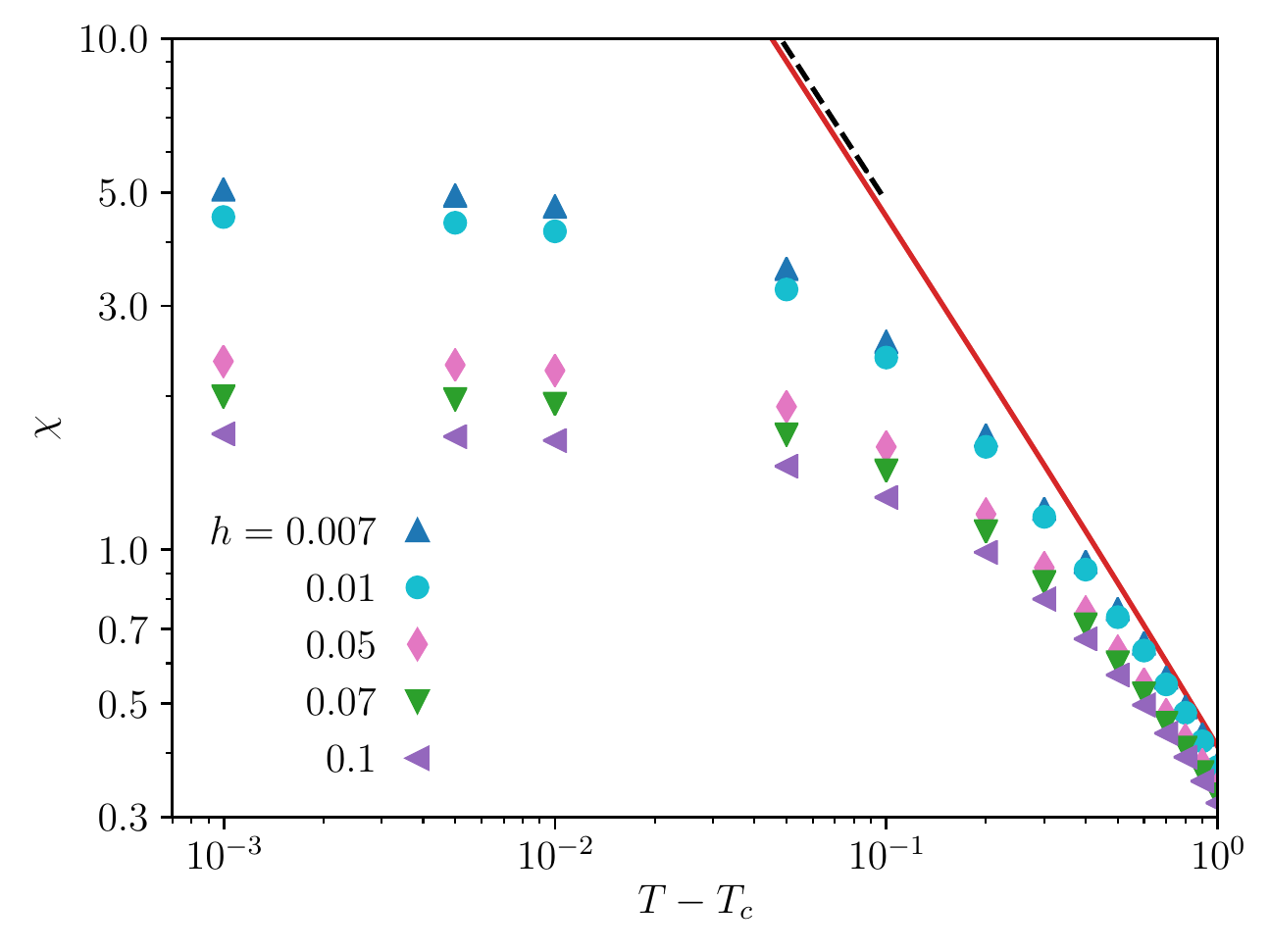}
 \includegraphics[width=7cm]{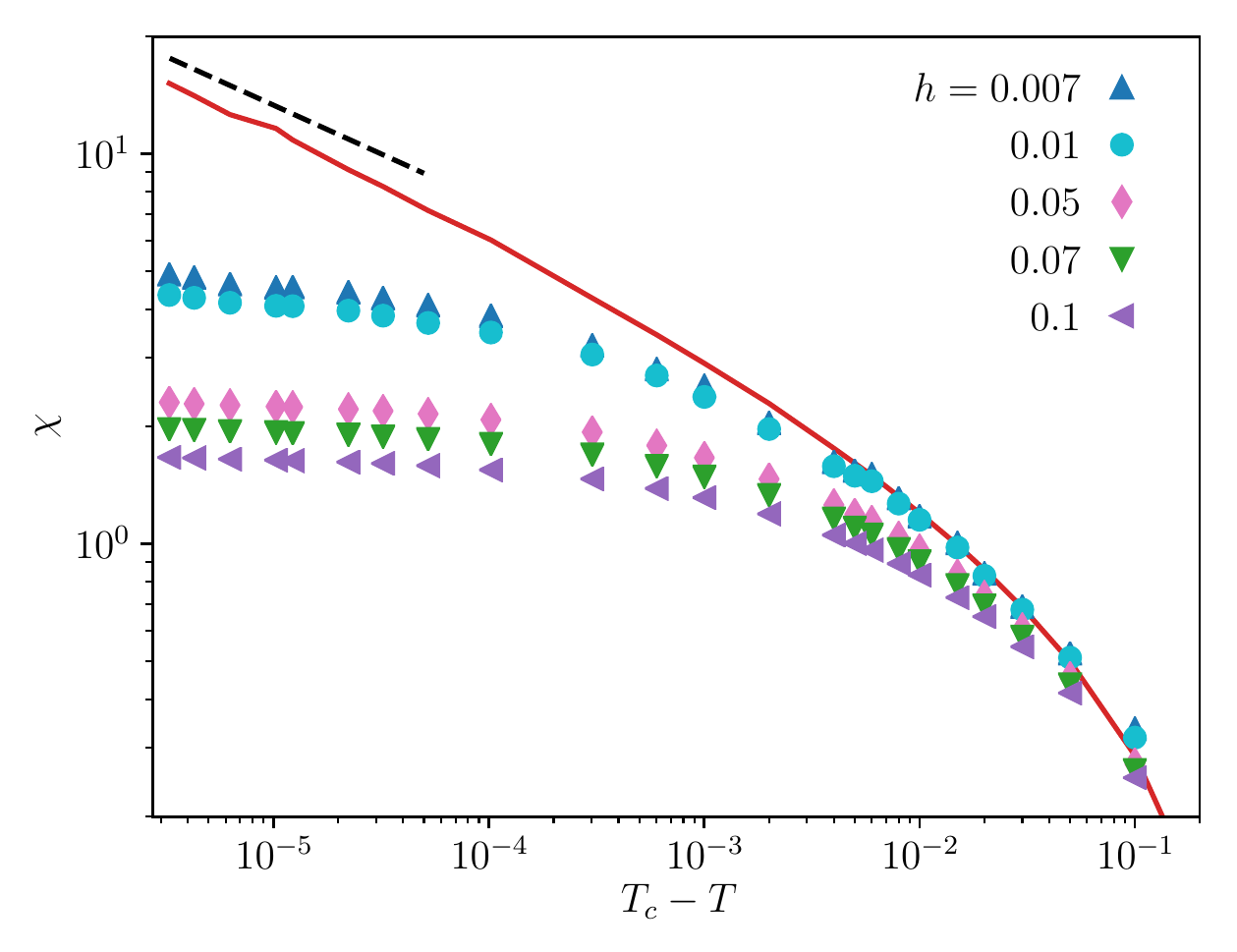}
  \caption{{\bf Model-$1$:} For $D=5$, the figure shows the susceptibility $\chi(T)$ denoted by points; here, the magnetization
  $m_{h}$ is obtained by numerically integrating the equations of
  motion~\eqref{eq:eom-Si} with $N=10^7$ and performing a time average
  of the instantaneous magnetization over an interval of length
  $20$, which is further averaged over $2$ realizations of the dynamics. The
      upper panel corresponds to the disordered phase $T>T_c\approx 0.476$,
      while the lower panel is for the ordered phase $T<T_c$. In
      either case, the black dashed line corresponds to the behavior
      close to $T_c$, Eq.~\eqref{eq:gamma}, with $\gamma^+=1$ and
      $\gamma^-=1/4$ as given by our
      theoretical predictions, see Table~\ref{tab:critical-exponents}.
      The red continuous lines in the figures are our theoretical
      result~\eqref{eq:chi}, with $L$ given by Eq.~\eqref{eq:LT} computed
      numerically by using the method detailed in Appendix~\ref{app:angle-average}.}
  \label{fig:n1-gamma}
\end{figure}

\begin{figure}[ht!]
  \centering
 \includegraphics[width=7cm]{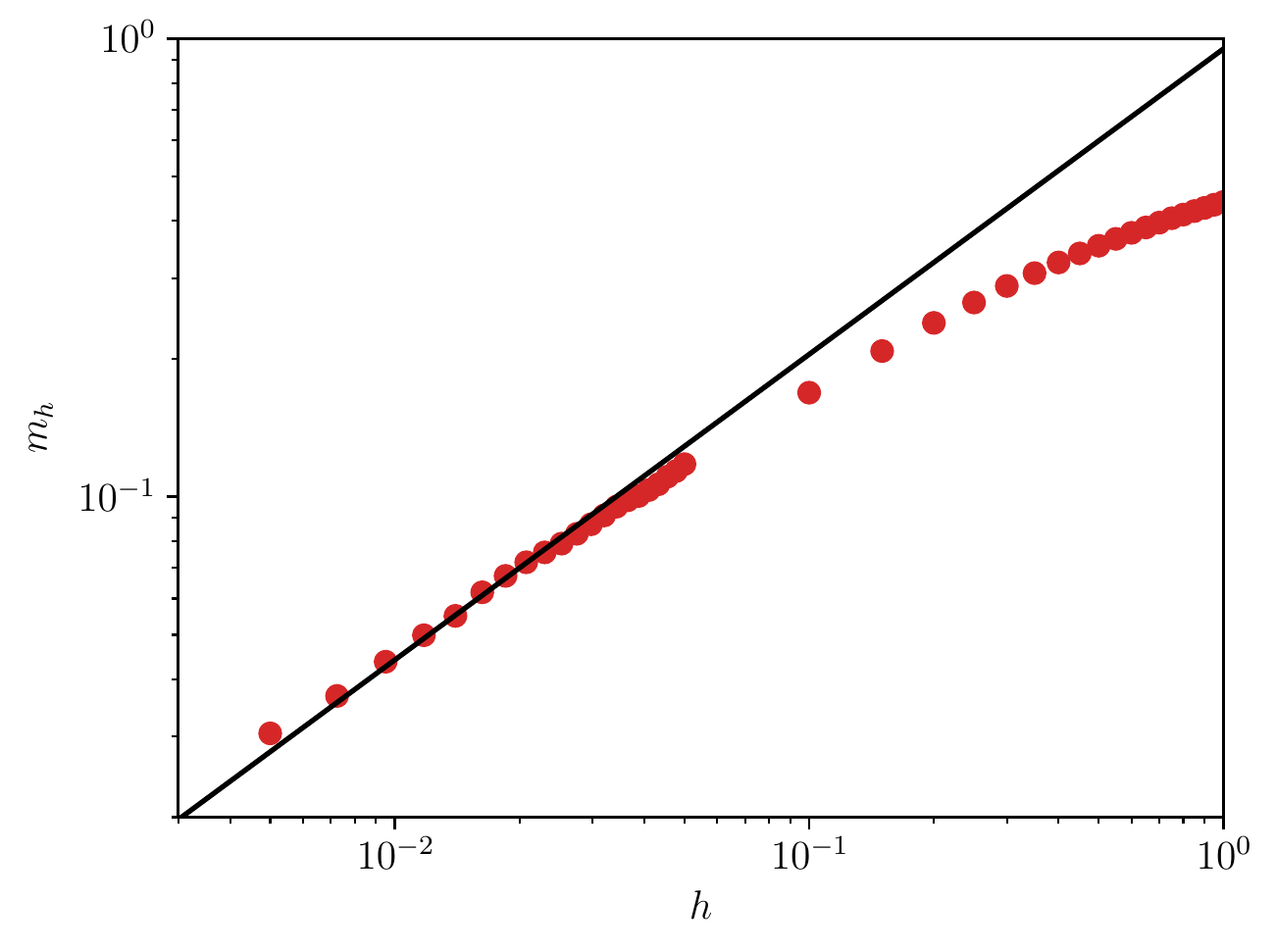}
  \caption{{\bf Model-$1$:} For $D=5$, the figure shows the nonlinear response at the critical point $T_c\approx 0.476$. The magnetization $m_{h}$, denoted by points, is obtained by
  numerically integrating the equations of motion~\eqref{eq:eom-Si} with
  $N=10^7$ and performing a time average of the instantaneous
  magnetization over an interval of length $20$, which is further averaged over $2$
  realizations of the dynamics. The black line corresponds to the behavior~\eqref{eq:delta}, with
  $\delta$ given by our theoretical analysis as $\delta=3/2$, see
  Table~\ref{tab:critical-exponents}. As expected, only for small $h$
  does our theory match with numerical results.}
  \label{fig:n1-delta}
\end{figure}

\begin{figure}[ht!]
  \centering
  \includegraphics[width=7cm]{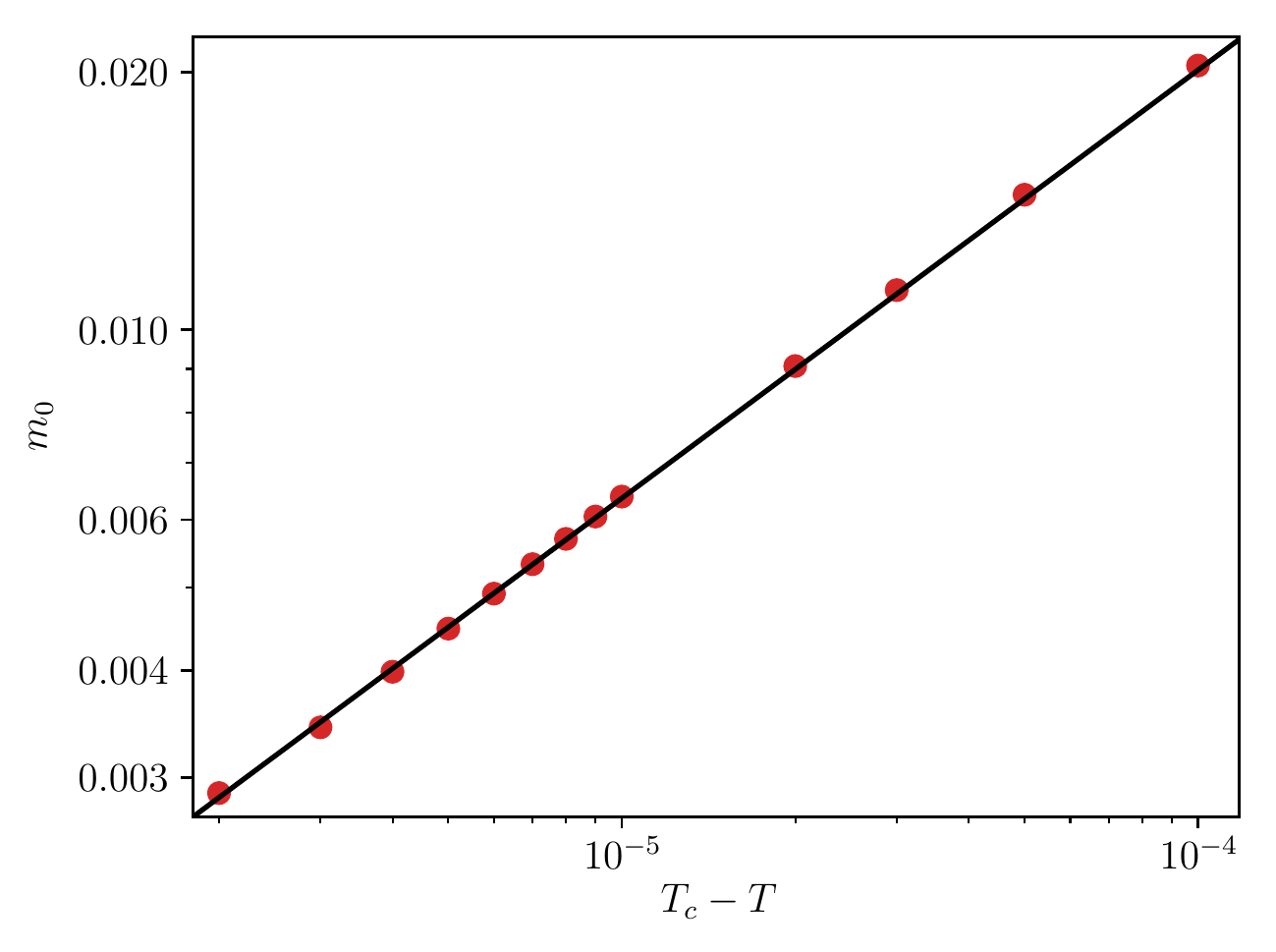}
  \caption{{\bf Model-$2$:} For $D=15$, the figure shows the spontaneous magnetization $m_{0}$ (points), obtained by
  solving the self-consistent equation \eqref{eq:self-consistent-m0} with thermal equilibrium as the reference state. The critical point is $T_c \approx 0.47$. 
  The line corresponds to the behavior \eqref{eq:beta}, with $\beta=1/2$, as
  predicted by our theory, see Table~\ref{tab:critical-exponents}.}
  \label{fig:n2-beta}
\end{figure}

\begin{figure}[ht!]
  \centering
\includegraphics[width=7cm]{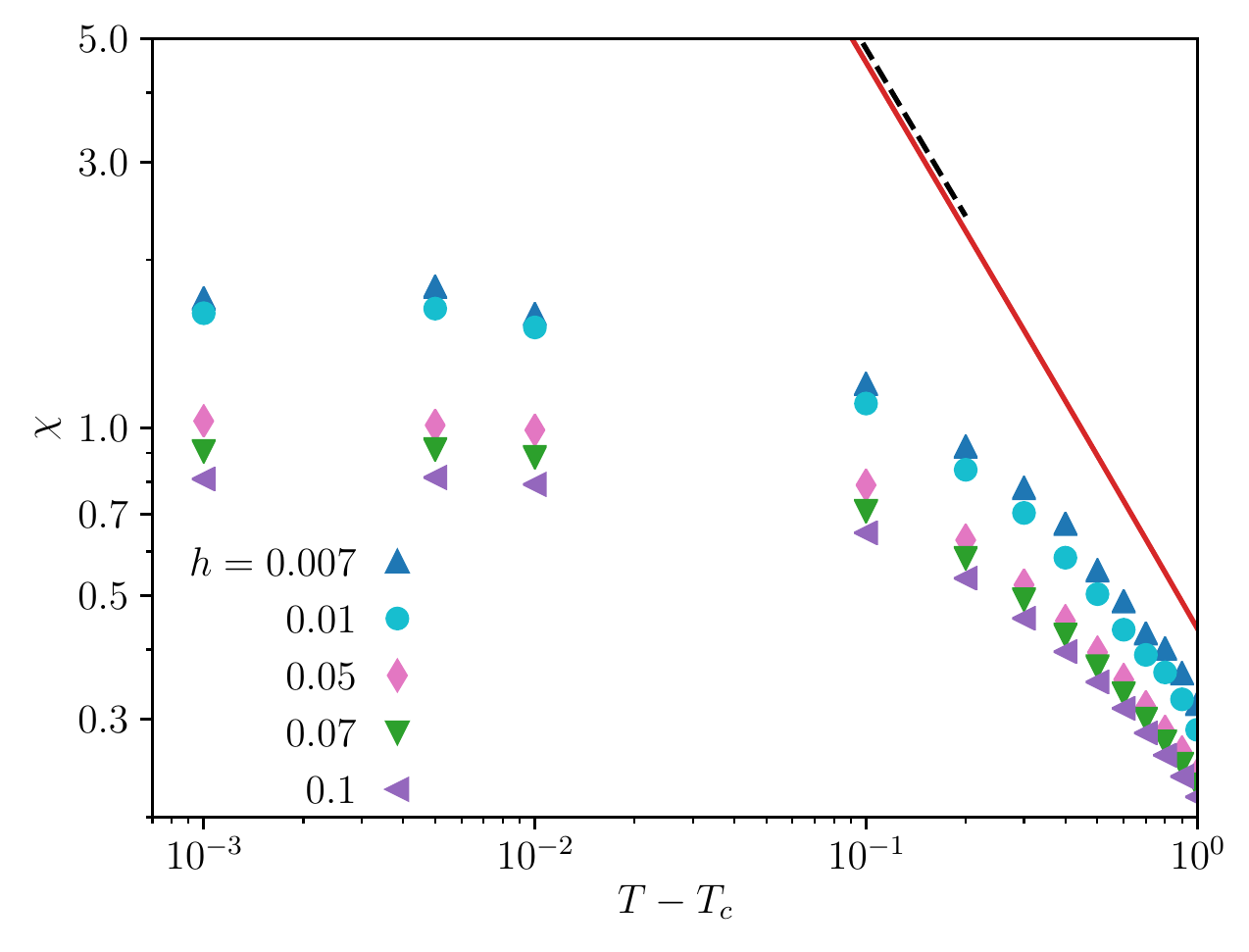}
 \includegraphics[width=7cm]{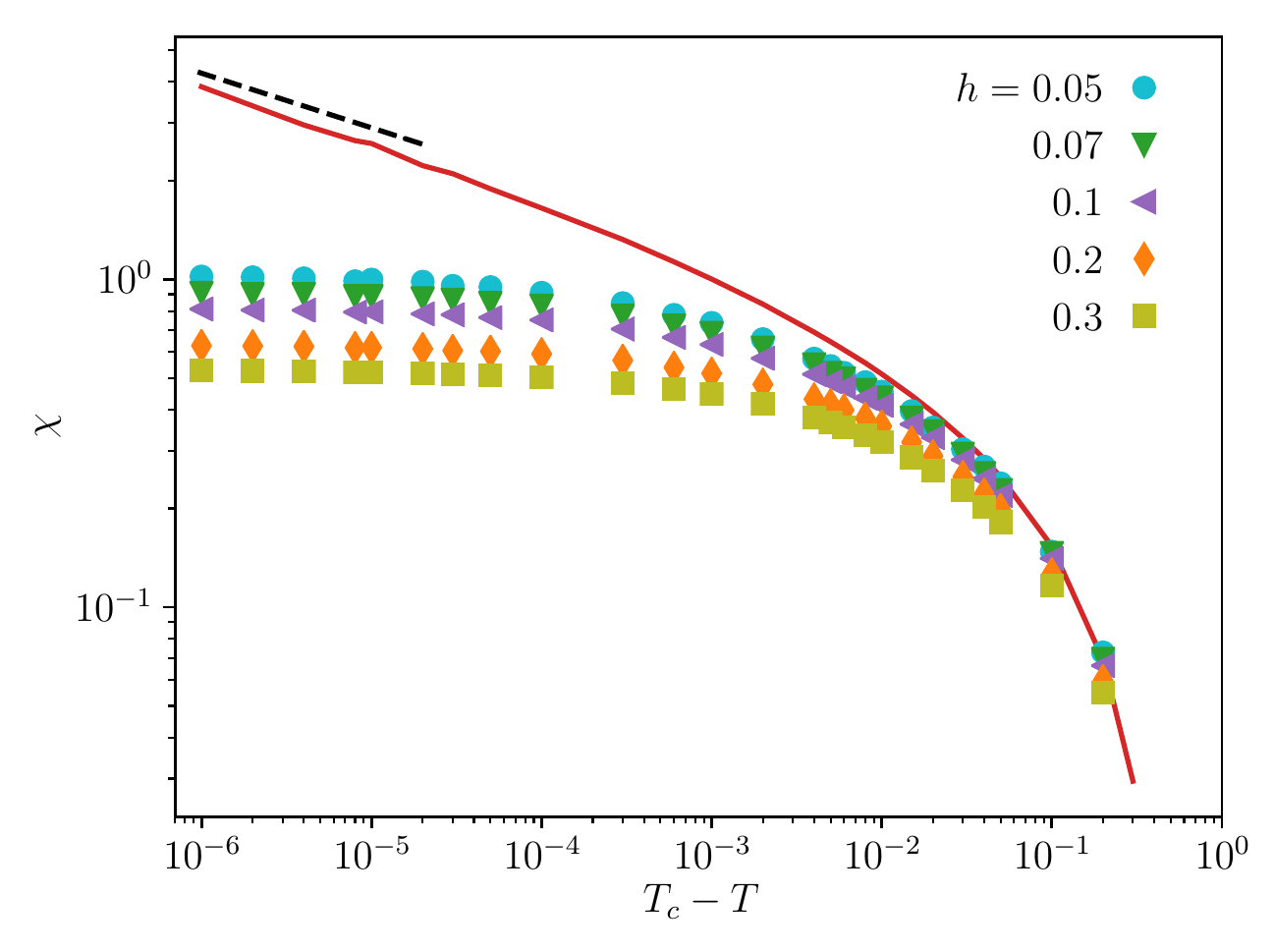}
  \caption{{\bf Model-$2$:} For $D=15$, the figure shows the susceptibility $\chi(T)$, denoted by points; here, the
  field-induced magnetization $m_{h}$ is obtained by numerically
  integrating the equations of motion~\eqref{eq:eom-Si} with $N=10^7$
  and performing a time average of the instantaneous magnetization over
  an interval of length $20$, which is further averaged over $2$
  realizations of the dynamics. The
      upper panel corresponds to the disordered phase $T>T_c \approx 0.47$,
      while the lower panel is for the ordered phase $T<T_c$. In
      either case, the black dashed line corresponds to the behavior
      close to $T_c$, Eq.~\eqref{eq:gamma}, with $\gamma^+=1$ and
      $\gamma^-=1/6$ as given by our
      theoretical predictions, see Table~\ref{tab:critical-exponents}.
      The red continuous lines in the figures are our theoretical
      result~\eqref{eq:chi}, with $L$ given by Eq.~\eqref{eq:LT} computed
      numerically by using the method given in Appendix~\ref{app:angle-average}.}
  \label{fig:n2-gamma}
\end{figure}

\begin{figure}[ht!]
  \centering
  \includegraphics[width=7cm]{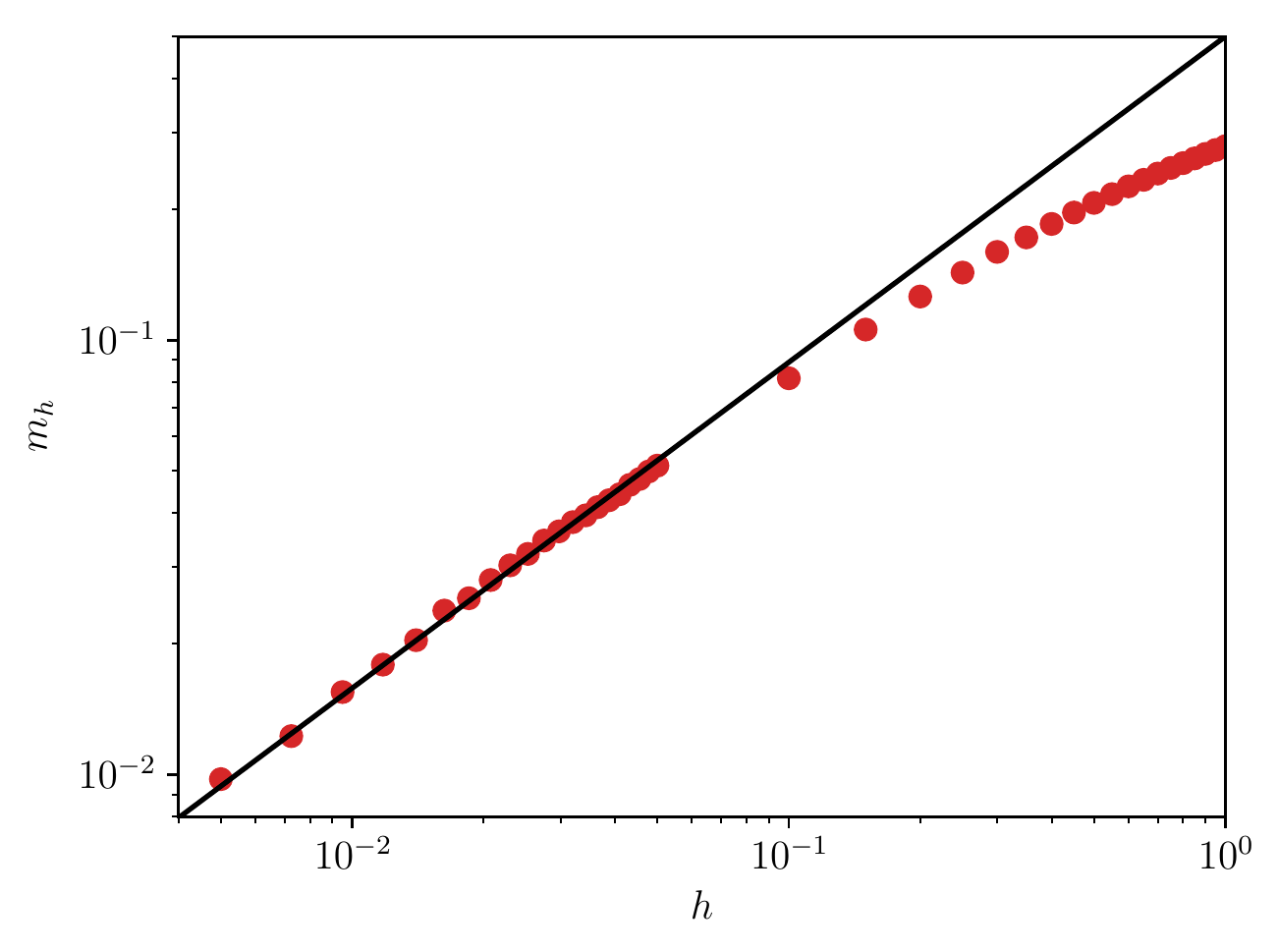}
  \caption{{\bf Model-$2$:} For $D=15$, the figure shows the nonlinear response at the critical point $T_c\approx 0.47$. The magnetization $m_{h}$, denoted by red points, is obtained by
  numerically integrating the equations of motion~\eqref{eq:eom-Si} with
  $N=10^7$ and performing a time average of the instantaneous
  magnetization over an interval of length $20$, which is further averaged over $2$
  realizations of the dynamics. The black line corresponds to the behavior~\eqref{eq:delta}, with
  $\delta$ given by our theoretical analysis as $\delta=4/3$, see
  Table~\ref{tab:critical-exponents}. As expected, only for small $h$
  does our theory match with numerical results.}
  \label{fig:n2-delta}
\end{figure}

Turning on a constant external field at $t=10$,
we study numerically the response of the system~\eqref{eq:H} with $N=10^{7}$.
Our results, presented in  Figs.~\ref{fig:n0-beta} and
\ref{fig:n0-gamma} for Model-$0$, Figs.~\ref{fig:n1-beta} --
\ref{fig:n1-delta} for Model-$1$, and in Figs.~\ref{fig:n2-beta} --
\ref{fig:n2-delta} for Model-$2$ are all consistent with our
theoretical predictions in Table~\ref{tab:critical-exponents}. For Model-$1$, we take $D=5$ for which $T_c \approx 0.476$ is obtained by using $G(x)=\exp(-x/T)$ in the expression given by Eq.~\eqref{eq:AT} for the quantity $A(T)$ and then solving $A(T_c)=0$, while for Model-$2$, we take $D=15$ for which one has $T_c \approx 0.47$. Note that
in Figs.~\ref{fig:n1-gamma} and \ref{fig:n2-gamma}, our theoretical
results match with our numerical results only for sufficiently small $h$, as
expected on the basis of the fact that our theoretical analysis is valid
in the linear response regime obtained in the limit $h \to 0$. Let us
remark that very close to $T_c$, numerical results for finite $N$ shown
in Figs.~\ref{fig:n1-gamma} and \ref{fig:n2-gamma} do not show the
divergence predicted by our theory and shown in these figures by red
lines, owing to finiteness of the field strength $h$ (the theoretical results are valid in the limit $h\to 0$ and $N \to \infty$ while satisfying the condition $h > 1/\sqrt{N}$ that ensures that the response dominates over finite-size fluctuations; in these figures, $N$ is large enough that the condition $h > 1/\sqrt{N}$ is satisfied, although not the limit $h\to 0$). Moreover, the convergence to the $h
\to 0$ limit is slower for Model-$2$ than for Model-$1$.

%%%%%%%%%%%%%%%%%%%%%%%%%%%%%%%%%%%%%%%%%%%%%%%%%%%%%%%%%%%%%%%%%%%%%%%%%%%%%%%%%%%%%%%%%%%
\section{Conclusions}
\label{sec:conclusions}

In this work, we have discussed response to an external field in mean-field systems of classical Heisenberg spins exhibiting a
second-order phase transition in the stationary state. The time evolution in the thermodynamic
limit of such systems is described by the so-called Vlasov equation for
the single-spin phase space distribution function. We have shown that
for Vlasov-stationary states that allow a second-order phase transition and when subject to a small external field, the critical exponents
characterizing power-law behavior of response close to the
critical point and obtained within the Vlasov dynamics may take values
different from the ones obtained on the basis of a statistical
mechanical analysis, with no reference to the dynamics of the initial
state in the presence
of the external field. Interestingly, we find
that both the sets of values of the critical exponents satisfy the same
scaling relation, which is incidentally the same as the one known for
mean-field Hamiltonian particle systems that are quite different from
the studied spin systems. This work hints on one hand at the universality
of critical behavior for mean-field systems evolving under Vlasov
dynamics, and cautions on the other hand against relying on a static
approach, with no reference to the dynamical
evolution, to extract critical exponent values for mean-field systems.

 The reason that one has to resort to the dynamics in order to
extract the correct critical exponents is the following. Let us first
recall the protocol we employ in extracting the critical exponents. We
prepare the system in Vlasov-stationary states parametrized by a
parameter $T$, and which allow a second-order phase transition and
consequently a critical point $T_{\rm c}$.
An example of such states is the thermal equilibrium
state for which the parameter $T$ is the temperature. We then subject
the system to a constant external field. As mentioned in the
introduction, mean-field systems like ours when considered in the
thermodynamic limit remain trapped in Vlasov-stationary states forever
in time. The dynamics in such states is non-ergodic due to existence of the Casimir invariants,
so that one may not apply tools from Boltzmann-Gibbs equilibrium statistical mechanics
to extract the critical exponents characterizing the response of such
states to the external field, for the simple reason that
ergodicity lies at the heart of the very foundation of equilibrium
statistical mechanics.

It may be noted that one could very well have a class of Vlasov-stationary states
that for our model do not
allow a second-order but a first-order phase transition. One may
mention the case of the HMF model where the Vlasov-stationary thermal
equilibrium state allows for a second-order phase transition, but there are
other classes of Vlasov-stationary states that allow a first-order
phase transition \cite{antoniazzi-07,Pakter:2011,rochafilho-amato-figueiredo-12,teles-benetti-pakter-levin-12}.
It would be interesting in the context of our model to study the response for other classes of Vlasov-stationary reference states that allow a first-order phase transition and differ from
the ones studied here. Another immediate follow up of this work
would be to investigate the validity, in the context of the studied spin
model, of the Kubo fluctuation-dissipation
theorem valid for short-range systems prepared in thermal equilibrium
and subject to small external fields. Studies in this direction are underway and comparison with the HMF model
results \cite{yamaguchi-16} will be reported elsewhere.
Collective $1/f$ fluctuation due to the Casimir invariants
is also an interesting topic to pursue \cite{yamaguchi-kaneko-18}.

%%%%%%%%%%%%%%%%%%%%%%%%%%%%%%%%%%%%%%%%%%%%%%%%%%%%%%%%%%%%%%%%%%%%%%%%%%%%%%%%%%%%%%%%%%%
\acknowledgments
The work of D.D. is supported by UGC-NET Research Fellowship Sr.
No. 2121450744, Ref. No. 21/12/2014(ii) EU-V.
Y.Y.Y. acknowledges the supports of JSPS KAKENHI Grant No. 16K05472.
The manuscript was finalized while S.G. was visiting the Quantitative 
Life Sciences section of the International Centre for Theoretical Physics (ICTP), 
Trieste, and he would
like to acknowledge the support and hospitality of the ICTP. S.G. acknowledges support from the Science and Engineering Research
Board (SERB), India, Grant No. TAR/2018/000023. 

%%%%%%%%%%%%%%%%%%%%%%%%%%%%%%%%%%%%%%%%%%%%%%%%%%%%%%%%%%%%%%%%%%%%%%%%%%%%%%%%%%%%%%%%%%%
\appendix

\section{Derivation of the critical exponents~\eqref{eq:exponents-SM}}
\label{sec:StatMech}

Here we discuss how one may obtain the values of the critical exponents
given in Eq.~\eqref{eq:exponents-SM}. The starting point is the
equivalent of Eqs.~\eqref{eq:f0}, \eqref{eq:H0} and
\eqref{eq:self-consistent} in presence of a constant field $\bh=(h,0,0)$:
\begin{equation}
  f_{h}(\phi,p) = F(H_{h}(\phi,p))
  = \dfrac{G(H_{h}(\phi,p))}{\iint_{\mu} G(H_{h}(\phi,p)) {\rm d}\phi
  {\rm d}p},
\end{equation}
with
\begin{equation}
  H_{h}(\phi,p) = Dp^{2n} - (\bbm_{h}+\bh)\cdot\bS,
\end{equation}
and
\begin{equation}
  \bbm_{h} = \iint_{\mu} \bS f_{h}(\phi,p) {\rm d}\phi {\rm d}p.
\end{equation}
The last equation gives
\begin{equation}
  m_{h}
  = \dfrac{\iint_{\mu} S_{x} G(Dp^{2n}-(m_{h}+h)S_{x}) {\rm d}\phi {\rm
  d}p}
  {\iint_{\mu} G(Dp^{2n}-(m_{h}+h) S_{x}) {\rm d}\phi {\rm d}p},
  \label{eq:mh}
\end{equation}
with $S_{x}=\sqrt{1-p^{2}}\cos\phi$.
We assume $G$ to be a smooth function of its argument. Since we take $G$
to be parametrized by the parameter $T$, this means that $G$ is also
smooth with respect to $T$. Let us expand $G$ in a Taylor series around
$Dp^{2n}$. Substituting the Taylor series and then integrating over
$\phi$, we see that in the numerator on the right hand side of Eq.~\eqref{eq:mh},
only odd order terms in $(m_h+h)$ survive; 
In the denominator, on the contrary,
only even order terms survive.
Consequently, the right hand side of Eq.~\eqref{eq:mh} has only odd
order terms of $m_{h}$, and the equation gives
\begin{equation}
  \label{eq:self-consistent-mh-SM}
  A (T)(m_{h}+h) + B(T) (m_{h}+h)^{3} - h = 0,
\end{equation}
with
\begin{equation}
  \label{eq:AT}
    A(T) \equiv 1 + \dfrac{1}{2}
    \dfrac{\int_{-1}^{1} (1-p^{2}) G'(Dp^{2n}) {\rm d}p}{\int_{-1}^{1}
    G(Dp^{2n}) {\rm d}p},
\end{equation}
and we have neglected the higher-order terms in $(m_h+h)$ in obtaining
Eq.~\eqref{eq:self-consistent-mh-SM}. Consider now the latter for
$h=0$. Assuming $B(T)>0$, one has a non-zero solution for $m_0$ for
$T<T_c$ and only a zero solution for $T>T_c$, where the critical point $T=T_c$ is where we have $A(T_c)=0$,
while $A(T)$ is positive (respectively, negative) for $T>T_c$
(respectively, $T<T_c$).
The nonzero solution, giving the spontaneous magnetization for $T<T_c$, is $m_{0}=\sqrt{-A/B}$.
The smoothness of $G$ with respect to the parameter $T$ allows $A(T)$ to
be expanded in a Taylor series around $T_c$, giving $A(T) \propto
(T-T_c)$ close to $T_c$, and this gives $\beta=1/2$.

The linear response is obtained by deriving
Eq.~\eqref{eq:self-consistent-mh-SM} with respect to $h$, 
and we have
\begin{equation}
  \chi(T) = \left. \dfrac{{\rm d}m_{h}}{{\rm d}h} \right|_{h=0}
  = \left\{
    \begin{array}{ll}
      (1-A)/A & (T>T_c), \\
      (1+2A)/(-2A) & (T<T_c). \\
    \end{array}
  \right.
\end{equation}
From the behavior $A(T) \propto (T-T_c)$ around $T=T_c$,
we get $\gamma^{\pm}=1$.

At the critical point, the self-consistent equation
\eqref{eq:self-consistent-mh-SM} reduces to
\begin{equation}
  B(T_c)(m_{h}+h)^{3} = h.
\end{equation}
The response $m_{h}\propto h^{1/\delta}~(\delta>1)$
is larger than $h$ for small $h$, so that we have $m_{h}^{3}\propto h$, implying $\delta=3$.

%%%%%%%%%%%%%%%%%%%%%%%%%%%%%%%%%%%%%%%%%%%%%%%%%%%%%%%%%%%%%%%%%%%%%%%%%%%%%%%%%%%%%%%%%%%
\section{Derivation of the response formula,
Eq.~\eqref{eq:response-formula}}
\label{app:response}
In this appendix, we summarize the derivation of the response formula \eqref{eq:response-formula}
by following Ref. \cite{ogawa-yamaguchi-14}.
Noting that the Vlasov equation is governed by the single-spin Hamiltonian
$H$, which depends on $f$ through the magnetization $\bbm$, the idea is to expand the Hamiltonian $H$ as
\begin{equation}
  H = H_{h} + K,
\end{equation}
where $H_{h}$ defined in \eqref{eq:Hh} is the asymptotic part,
characterizing the stationary ($t \to \infty$) state $f_h$, while $K$ is the transient part.
For our spin model, the explicit form of the transient part is
\begin{equation}
  K(\phi,p,t) = - m_{\rm T}(t) \sqrt{1-p^{2}}\cos\phi,
\end{equation}
where the transient magnetization $m_{\rm T}(t)$ is obtained as
\begin{equation}
  m_{\rm T}(t) = \iint_{\mu} \sqrt{1-p^{2}}\cos\phi~ g(\phi,p) {\rm
  d}\phi {\rm d}p,
\end{equation}
and the transient state $g$ is defined by $g=f-f_{h}$.
The transient quantities, $g,K$ and $m_{\rm T}(t)$, are not known a
priori,
but they do not appear in the final result of the response formula.

Let us write the Vlasov equation \eqref{eq:Vlasov} as
\begin{equation}
  \label{eq:Vlasov-L} 
  \dfracp{f}{t} = \mathcal{L}_{H}f
  = \mathcal{L}_{H_{h}}f + \mathcal{L}_{K}f,
\end{equation}
where the linear operator $\mathcal{L}_{H}$ is defined as
\begin{equation}
  \mathcal{L}_{H}f
  \equiv \dfracp{H}{\phi} \dfracp{f}{p} - \dfracp{H}{p} \dfracp{f}{\phi}. 
\end{equation}
Eq.~\eqref{eq:Vlasov-L} is still exact.
Now, we assume that contribution from the transient part,
$\mathcal{L}_{K}f$, is negligible, which is justified under
some assumptions for Hamiltonian systems and
may be related to the phenomenon of Landau damping, see Ref.~\cite{ogawa-yamaguchi-14} for details.

Under the aforementioned assumption, the formal solution
to the Vlasov equation \eqref{eq:Vlasov-L} is 
\begin{equation}
  f(\phi,p,t) = \exp[t\mathcal{L}_{h}]f_{0}(\phi,p),
\end{equation}
which represents temporal evolution of $f_{0}$ under Hamiltonian flow associated with the asymptotic Hamiltonian $H_{h}$.
Assuming ergodicity, a formula that replaces the time average with a
partial phase-space average with respect to a iso-$H_{h}$ surface gives
\begin{equation}
  \lim_{t\to\infty} \dfrac{1}{t} \int_{0}^{t} e^{s\mathcal{L}_{H}}
  f_{0}(\phi,p) {\rm d}s
  = \ave{f_{0}}_{h}.
\end{equation}
Noting that the left hand side is nothing but the asymptotic stationary state $f_{h}$, we obtain the response formula \eqref{eq:response-formula}.

%%%%%%%%%%%%%%%%%%%%%%%%%%%%%%%%%%%%%%%%%%%%%%%%%%%%%%%%%%%%%%%%%%%%%%%%%%%%%%%%%%%%%%%%%%%
\section{Derivation of the estimation \eqref{eq:L2-estimation} for the quantity $L_{2}$}
\label{sec:estimation-L2}

Here, we derive Eq.~\eqref{eq:L2-estimation}. For simplicity of notation, we use the same symbols $(w,I)$ for angle-action variables associated
with the single-spin Hamiltonian $H_{0}$ as the ones used for $H_{h}$ in Sec. \ref{sec:response-formula}, 
but the latter do not appear in this section and no confusion should
arise. Here we consider positive $n$.

We start with Eq.~\eqref{eq:L2}. 
Noting that $\ave{S_{x}}_{0}^{2}$ and $H_{0}$
depend on only the action variable $I$,
we rewrite $L_{2}$ as
\begin{equation}
  L_{2} = \iint_{\mu} \dfrac{\psi(I)}{2\pi} {\rm d}w {\rm d}I
  = \int \psi(I) {\rm d}I,
\end{equation}
where
\begin{equation}
  \psi(I) = - 2\pi \ave{S_{x}}_{0}^{2}F'(H_{0}(I)),
\end{equation}
and we have used the fact that canonical transformation from $(\phi,p)$ to $(w,I)$ gives
${\rm d}\phi {\rm d}p={\rm d}w{\rm d}I$.
The single-spin Hamiltonian $H_{0}$ has a separatrix
which consists of the stable and unstable manifolds of the fixed point
$(S_{x},S_{y},S_{z})=(-1,0,0)$ and encloses the point $O=(1,0,0)$ on the phase space of the unit sphere.
On the basis of this observation, we make an essential assumption for
estimating $L_{2}$, namely, that the main contribution to $L_{2}$ comes
from the region around the point $O$. We now change twice the variables
of integration in $L_{2}$. First, to divide the phase space into the
inside and the outside
of the separatrix, the integration variable $I$ is changed to energy $E$ as
\begin{equation}
  \label{eq:L2-E}
  L_{2} = \int_{E_{\rm min}}^{E_{\rm sep}} \dfrac{\psi(I)}{\Omega(I)}
  {\rm d}E
  + \int_{E_{\rm sep}}^{E_{\rm max}} \dfrac{\psi(I)}{\Omega(I)} {\rm d}E,
\end{equation}
where the frequency $\Omega(I)$ is defined by
\begin{equation}
  \Omega(I) \equiv \dfrac{{\rm d}H_{0}}{{\rm d}I}(I).
\end{equation}
Let $E_{\rm min},E_{\rm sep}$ and $E_{\rm max}$ denote respectively
the minimum energy, the separatrix energy, and the maximum energy.
Following the essential assumption, we omit the second term of $L_{2}$
in \eqref{eq:L2-E}.
Second, to eliminate the dependence on $m_{0}$ of the integration interval,
$E$ is changed to a variable $k$ defined as
\begin{equation}
  k \equiv \dfrac{E - E_{\min}}{E_{\rm sep}-E_{\rm min}}.
\end{equation}
Consequently, one has
\begin{equation}
  L_{2} \simeq \int_{0}^{1} \psi(I(k))
  \dfrac{E_{\rm sep}-E_{\rm min}}{\Omega(I(k))} {\rm d}k
  = 2m_{0} \int_{0}^{1} \dfrac{\psi(k)}{\Omega(k)} {\rm d}k,
\end{equation}
where $\psi(I(k))$ is simply denoted as $\psi(k)$ for instance.

To estimate $\Omega$ around the point $O$, which corresponds to $(\phi,p)=(0,0)$, we approximate $H_{0}$ as
\begin{equation}
  H_{0}(\phi,p) \simeq Dp^{2n} + \dfrac{m_{0}}{2}\phi^{2}.
\end{equation}
The action variable is the area enclosed by a periodic orbit,
and hence, we have
\begin{equation}
  I = \dfrac{1}{2\pi} \oint \phi {\rm d}p
  = \dfrac{2}{\pi} \sqrt{\dfrac{2}{m_{0}}}
  \int_{0}^{p_{\rm max}} \sqrt{E-Dp^{2n}} {\rm d}p,
\end{equation}
with $p_{\rm max}=(E/D)^{1/(2n)}$.
In terms of 
\begin{equation}
  p \equiv \left( \dfrac{E}{D} \right)^{1/(2n)} u,
\end{equation}
we have the action variable as
\begin{equation}
  I = I_{0} \dfrac{E^{(n+1)/2n}}{\sqrt{m_{0}}},
  \quad
  I_{0} = \dfrac{2\sqrt{2}}{\pi D^{1/(2n)}} \int_{0}^{1}
  \sqrt{1-u^{2n}}  {\rm d}u,
\end{equation}
which gives
\begin{equation}
  E = m_{0}^{n/(n+1)} \left( \dfrac{I}{I_{0}} \right)^{2n/(n+1)}.
\end{equation}
The frequency $\Omega$ is therefore
\begin{equation}
  \Omega = m_{0}^{n/(n+1)} \widetilde{\Omega},
  \quad 
  \widetilde{\Omega} = \dfrac{2n}{n+1} \dfrac{1}{I_{0}} \left( \dfrac{I}{I_{0}} \right)^{(n-1)/(n+1)}.
\end{equation}
Putting all together, we have the estimation of $L_{2}$ as
\begin{equation}
\label{eq:L2-app}
  L_{2} \simeq 2 m_{0}^{1/(n+1)} \int_{0}^{1}
  \dfrac{\psi(k)}{\widetilde{\Omega}(k)} {\rm d}k.
\end{equation}
We remark that $\ave{S_{x}}_{0}$ is zero in the disordered phase, but it
does not vanish in the ordered phase even when the limit $m_{0}\to 0$ is taken,
because the iso-$H_{0}$ line is confined to the direction $\phi$
around the point $O$ corresponding to $(\phi,p)=(0,0)$.
Equation~\eqref{eq:L2-app} yields the estimation
\eqref{eq:L2-estimation} of the main text. 

%%%%%%%%%%%%%%%%%%%%%%%%%%%%%%%%%%%%%%%%%%%%%%%%%%%%%%%%%%%%%%%%%%%%%%%%%%%%%%%%%%%%%%%%%%%
\section{A method to compute averages $\ave{\cdot}_{0}$ over angles}
\label{app:angle-average}
In this appendix, we discuss a method to compute the angle average $\ave{\cdot}_{0}$.
The single-spin Hamiltonian $H_{0}$ has the angle-action variables $(w,I)$
whose temporal evolution is
\begin{equation}
  w(t) = w(0) + \Omega(I)t,
  \quad
  I(t) = I(0), 
\end{equation}
where $\Omega(I)={\rm d}H_{0}/{\rm d}I$.
Changing the variable from $w$ to $t$, we have the average as
\begin{equation}
  \begin{split}
    \ave{B}_{0}
    & = \dfrac{\int_{0}^{2\pi} B(w,I) {\rm d}w}{\int_{0}^{2\pi} {\rm d}w}
    = \dfrac{\int_{0}^{T_{p}} B(w(t),I) \Omega(I) {\rm
    d}t}{\int_{0}^{T_{p}} \Omega(I) {\rm d}t} \\
    & = \dfrac{\int_{0}^{T_{p}} B(w(t),I) {\rm d}t}{\int_{0}^{T_{p}}
    {\rm d}t}, 
  \end{split}
\end{equation}
where $T_{p}$ is the period on the considered iso-$I$ line.
We may write down the canonical equations of motion for
$H_{0}=Dp^{2n}-m_{0}\sqrt{1-p^{2}}\cos\phi$ as
\begin{equation}
  \begin{split}
    & \dfrac{{\rm d}\phi}{{\rm d}t} = 2nDp^{2n-1} + m_{0} \dfrac{p}{\sqrt{1-p^{2}}} \cos\phi, \\
    & \dfrac{{\rm d}p}{{\rm d}t} = - m_{0} \sqrt{1-p^{2}} \sin\phi,
  \end{split}
\end{equation}
and the average $\ave{\sqrt{1-p^{2}}\cos\phi}_{0}$ is computed as
\begin{equation}
  \ave{\sqrt{1-p^{2}}\cos\phi}_{0}
  = \dfrac{\int_{0}^{T_{p}} \sqrt{1-p^{2}(t)} \cos\phi(t)~{\rm d}t}{\int_{0}^{T_{p}} {\rm d}t}.
\end{equation}
Note that the left hand side depends on the action $I$ only, with the
initial condition $(\phi(0),p(0))$ determining the value of $I$.

%%%%%%%%%%%%%%%%%%%%%%%%%%%%%%%%%%%%%%%%%%%%%%%%%%%%%%%%%%%%%%%%%%%%%%%%%%%%%%%%%%%%%%%%%%%%%%%%%%%
%%%%%%%%%%%%%%%%%%%%%%%%%%%%%%%%%%%%%%%%%%%%%%%%%%%%%%%%%%%%%%%%%%%%%%%%%%%%%%%%%%%%%%%%%%%%%%%%%%%

%%%%%%%%%%%%%%%%%%%%%%%%%%%%%%%%%%%%%%%%%%%%%%%%%%%%%%%%%%%%%%%%%%%%%%%%%%%%%%%%%%%%%%%%%%%%%%%%%%%

\begin{thebibliography}{99}
\bibitem{Fisher}M. E. Fisher, The theory of equilibrium critical
phenomena, Rep. Prog. Phys. {\bf 30}, 615 (1967).

\bibitem{Stanley}H. E. Stanley, {\it Introduction to Phase Transitions
and Critical Phenomena} (Oxford University Press, UK, 1987).

\bibitem{nishimori-ortiz-11}
  H. Nishimori and G. Ortiz,
  {\it Elements of Phase Transitions and Critical Phenomena}
  (Oxford University Press, Oxford, 2011).
  
\bibitem{campa-dauxois-ruffo-09}
  A. Campa, T. Dauxois and S. Ruffo,
  Statistical mechanics and dynamics of solvable models with long-range interactions,
  Phys. Rep. {\bf 480}, 57 (2009).


\bibitem{levin-etal-14}
  Y. Levin, R. Pakter, F. B. Rizzato, T. N. Teles and F. P. C. Benetti,
  Nonequilibrium statistical mechanics of systems with long-range interactions,
  Phys. Rep. {\bf 535}, 1 (2014).

\bibitem{Campa:2014}
  A. Campa, T. Dauxois, D. Fanelli and S. Ruffo, 
  {\it Physics of Long-range Interacting Systems}
  (Oxford University Press, Oxford, 2014).

\bibitem{Gupta:2017}S. Gupta and S. Ruffo, The world of long-range
interactions: A bird’s eye view, Int. J. Mod. Phys. A {\bf
32}, 1741018 (2017). 

\bibitem{yamaguchi-etal-04}
  Y. Y. Yamaguchi, J. Barr{\'e}, F. Bouchet, T. Dauxois and S. Ruffo,
  Stability criteria of the Vlasov equation and quasi-stationary states of the HMF model,
  Physica A {\bf 337}, 36 (2004).

\bibitem{binney-tremaine-08}
  J. Binney and S. Tremaine,
  {\it Galactic Dynamics, 2nd ed.}
  (Princeton University Press, Princeton, NJ, 2008).


\bibitem{barre-etal-06}
  J. Barr{\'e}, F. Bouchet, T. Dauxois, S. Ruffo, and Y. Y. Yamaguchi,
  The Vlasov equation and the Hamiltonian mean-field model,
  Physica A {\bf 365}, 177 (2006).
  
\bibitem{antoniazzi-07}
  A. Antoniazzi, D. Fanelli, S. Ruffo, and Y. Y. Yamaguchi,
  Nonequilibrium Tricritical Point in a System with Long-Range Interactions,
  Phys. Rev. Lett. {\bf 99}, 040601 (2007).

\bibitem{Pakter:2011}
  R. Pakter and Y. Levin, Core halo distribution in
  the Hamiltonian mean-field model, Phys. Rev. Lett. {\bf 106}, 200603 (2011).

\bibitem{rochafilho-amato-figueiredo-12}
  T. M. Rocha Filho, M. A. Amato, and A. Figueiredo,
  Nonequilibrium phase transitions and violent relaxation in the Hamiltonian mean-field model,
  Phys. Rev. E {\bf 85}, 062103 (2012).

\bibitem{teles-benetti-pakter-levin-12}
   T. N. Teles, F. P. da C. Benetti, R. Pakter, and Y. Levin,
   Nonequilibrium Phase Transitions in Systems with Long-Range Interactions,
   Phys. Rev. Lett. {\bf 109}, 230601 (2012).

\bibitem{ogawa-patelli-yamaguchi-14}
  S. Ogawa, A. Patelli and Y. Y. Yamaguchi,
  Non-mean-field critical exponent in a mean-field model: Dynamics versus statistical mechanics,
  Phys. Rev. E {\bf 89}, 032131 (2014).

\bibitem{inagaki-konishi-93}
  S. Inagaki and T. Konishi,
  Dynamical Stability of a Simple Model Similar to Self-Gravitating Systems,
  Publ. Astron. Soc. Japan {\bf 45}, 733 (1993).

\bibitem{antoni-ruffo-95}
  M. Antoni and S. Ruffo,
  Clustering and relaxation in Hamiltonian long-range dynamics,
  Phys. Rev. E {\bf 52}, 2361 (1995).

\bibitem{ogawa-yamaguchi-15}
  S. Ogawa and Y. Y. Yamaguchi,
  Landau-like theory for universality of critical exponents in quasistationary states of isolated mean-field systems,
  Phys. Rev. E {\bf 91}, 062108 (2015).

\bibitem{mazur-69}
  P. Mazur,
  Non-ergodicity of phase functions in certain systems,
  Physica (Amsterdam) {\bf 43}, 533 (1969).

\bibitem{suzuki-71}
  M. Suzuki,
  Ergodicity, constants of motion, and bounds for susceptibilities,
  Physica (Amsterdam) {\bf 51}, 277 (1971).

\bibitem{Kac:1963}M. Kac, G. E. Uhlenbeck and P. C. Hemmer, On the van
der Waals Theory of the Vapor-Liquid Equilibrium. I. Discussion of a
one-dimensional model, J. Math.
Phys. {\bf 4}, 216 (1963).

\bibitem{Gupta:2011}
  S. Gupta and D. Mukamel, Quasistationarity in a model of
  classical spins with long-range interactions,
  J. Stat. Mech.: Theory Exp., P03015 (2011).

\bibitem{Barre:2014}J. Barr\'{e} and S. Gupta, Classical Heisenberg spins with long-range interactions: relaxation to equilibrium for finite
systems, J. Stat. Mech.: Theory
Exp. P02017 (2014).

\bibitem{kventsel-katriel-84}
  G. F. Kventsel and J. Katriel,
  Static properties of infinite-range spin Hamiltonians,
  Phys. Rev. B {\bf 30}, 2828 (1984).

\bibitem{ogawa-yamaguchi-12}
  S. Ogawa and Y. Y. Yamaguchi,
  Linear response theory in the Vlasov equation for homogeneous and for inhomogeneous quasistationary states,
  Phys. Rev. E {\bf 85}, 061115 (2012).

\bibitem{note}In this work, we measure temperature in units of the Boltzmann
constant. 

\bibitem{Pakter:2013}
  R. Pakter and Y. Levin, Nonequilibrium dynamics of an infinite range
  XY model in an external field, J. Stat. Phys. {\bf 150}, 531 (2013).

  
\bibitem{ogawa-yamaguchi-14}
  S. Ogawa and Y. Y. Yamaguchi,
  Nonlinear response for external field and perturbation in the Vlasov system,
  Phys. Rev. E {\bf 89}, 052114 (2014).

\bibitem{yamaguchi-16}
  Y. Y. Yamaguchi,
  Strange scaling and relaxation of finite-size fluctuation in thermal equilibrium,
  Phys. Rev. E {\bf 94}, 012133 (2016).

\bibitem{yamaguchi-kaneko-18}
  Y. Y. Yamaguchi and K. Kaneko,
  Collective $1/f$ fluctuation by pseudo-Casimir-invariants,
  Phys. Rev. E {\bf 98}, 020201(R) (2018).
  
\end{thebibliography}
\end{document}